\documentstyle[12pt]{article}
\begin{document}
\begin{center}
{\bf \LARGE Generalized Chiral $QED_2$ :\\
Anomaly and Exotic Statistics}

\vspace{2 cm}

{\bf \Large Fuad M. Saradzhev}

\vspace{1 cm}

{\small \it Institute of Physics, Academy of Sciences of Azerbaijan,\\
\it Huseyn Javid pr. 33, 370143 Baku, AZERBAIJAN}\\

\end{center}

\vspace{3 cm}

\begin{flushleft}

{\bf ABSTRACT}

\end{flushleft}

\rm

We study the influence of the anomaly on the physical quantum
picture of the generalized chiral Schwinger model defined on $S^1$.
We show that the anomaly  i) results in the background linearly
rising electric field and ii) makes the spectrum of the
physical Hamiltonian nonrelativistic without a massive boson.
The physical matter fields acquire exotic statistics.
We construct explicitly the algebra  of the Poincare generators 
and show that it differs from the Poincare one. We exhibit the role 
of the vacuum Berry phase 
in the failure of the Poincare algebra to close. We prove that, in 
spite of 
the background electric field, such phenomenon as the total screening
of external charges characteristic for the standard Schwinger 
model takes place in the generalized chiral Schwinger 
model, too.

\vspace{1 cm}


\newpage
\section{Introduction}
\label{sec: intro}
\rm
  
The two-dimensional $QED$ with massless fermions, i.e. the Schwinger
model (SM), demonstrates such phenomena as the dynamical mass
generation and the total screening of the charge \cite{schw63} .
Although the Lagrangian of the SM contains only massless fields, a
massive boson field emerges out of the interplay of the dynamics
that govern the original fields. This mass generation is due to the
complete compensation of any external charge inserted into the vacuum.

In the chiral Schwinger model (CSM) \cite{jack85,raja85}
the right and left chiral components of the fermionic field have 
different charges. The left-right asymmetric
matter content leads to an anomaly. At the quantum level, the local
gauge symmetry is not realized by a unitary action of the gauge 
symmetry group on Hilbert space. The Hilbert space furnishes a
projective representation of the symmetry group 
\cite{wign39,jack83,nels85}.

In this paper, we aim to study the influence of the anomaly on the
physical quantum picture of the CSM. Do the dynamical mass generation
and the total screening of charges take place also in the CSM? Are
there any new physical effects caused just by the left-right 
asymmetry?
These are the questions which we want to answer.

To get the physical quantum picture of the CSM we need first to
construct a self-consistent quantum theory of the model and then 
solve all the quantum constraints. In the quantization procedure,
the anomaly manifests itself through a special Schwinger term in the
commutator algebra of the Gauss law generators. This term changes
the nature of the Gauss law constraint: instead of being first-class
constraint, it turns into second-class one. As a consequence, the
physical quantum states cannot be defined as annihilated by the
Gauss law generator.

There are different approaches to overcome this problem and to
consistently quantize the CSM. The fact that the second class
constraint appears only after quantization means that the number
of degrees of freedom of the quantum theory is larger than that of
the classical theory. To keep the Gauss law constraint first-class,
Faddeev and Shatashvili proposed adding an auxiliary field in such a
way that the dynamical content of the model does not change
\cite{fadd86}. At the same time, after quantization it is the 
auxiliary
field that furnishes the additional "irrelevant" quantum degrees of
freedom. The auxiliary field is described by the Wess-Zumino term.
When this term is added to the Lagrangian of the original model,
a new, anomaly-free model is obtained. Subsequent canonical
quantization of the new model is achieved by the Dirac procedure.

For the CSM, the correspondig WZ-term is not defined uniquely.
It contains the so called Jackiw-Rajaraman parameter $a > 1$.
This parameter reflects an ambiguity in the bosonization procedure
and in the construction of the WZ-term. The spectrum of
the new, anomaly-free model turns out to be relativistic and
contains a relativistic boson. However, the mass of the boson also 
depends
on the Jackiw-Rajaraman parameter \cite{jack85,raja85}. This
mass corresponds therefore to the 
"irrelevant" quantum degrees of freedom. The quantum theory 
with such a parameter
in the spectrum is not physical ,
i.e. that final
version of the quantum theory which we would like to get.
The latter should not contain any nonphysical parameters , 
otherwise one can not  say anything  
about a physical  quantum picture.

In another approach also formulated by Faddeev \cite{fadd84},
the auxiliary field is not added, so the quantum Gauss law constraint
remains second-class. The standard Gauss law is assumed to be regained
as a statement valid in matrix elements between some states of the
total Hilbert space, and it is the states that are called physical.
The theory is regularized in such a way that the quantum Hamiltonian
commutes with the nonmodified, i.e. second-class quantum Gauss law
constraint. The spectrum turns out to be  
non-relativistic \cite{hall86,para88}.

Here, we follow the approach given in our previous work 
\cite{sarad91}. The pecularity of the CSM is that its anomalous
behaviour is trivial in the sense that the second class constraint
which appears after quantization can be turned into first class by
a simple redefinition of the canonical variables. This allows us
to formulate a modified Gauss law to constrain physical states.
The physical states are gauge-invariant up to a phase,
the phase being $1$-cocycle of the gauge symmetry group algebra. 
In \cite{niemi85,niemi86,semen87}, the modification of the Gauss
law constraint is obtained by making use of the adiabatic approach.

Contrary to \cite{sarad91} where the CSM is defined
on $R^1$ , we suppose here that space is a circle of length $L$,
$-\frac{L}{2} \leq x < \frac{L}{2}$ , so space-time manifold is a 
cylinder
$S^1 \times R^1$ . The gauge field then acquires a global physical
degree of freedom represented by the non-integrable phase of the
Wilson integral on $S^1$. We show that this brings in the physical
quantum picture new features of principle.

Another way of making two-dimensional gauge field dynamics nontrivial
is by fixing the spatial asymptotics of the gauge field 
\cite{sarad92,sarad94}. If we assume that the gauge field defined on
$R^1$ diminishes rather rapidly at spatial infinities, then it again
acquires a global physical degree of freedom. We will see that the
physical quantum picture for the model defined on $S^1$ is equivalent
to that obtained in \cite{sarad92,sarad94}.

We consider the general version of the CSM with a $U(1)$ gauge
field coupled with different charges to both chiral components
of a fermionic field. We show that the charges are not arbitrary,
but satisfy a quantization condition. The SM where these charges 
are equal is a special case of the generalized CSM. This will allow
us at each step of our consideration to see the distinction between
the two models.

We work in the temporal gauge $A_0=0$ in the framework of the
canonical quantization scheme and the Dirac's quantization method
for the constrained systems \cite{dirac64}. We use the system of
units where $c=1$. In Section~\ref{sec: quant}, we quantize our 
model in two steps. First, the matter fields
are quantized, while $A_1$ is handled as a classical background field.
The gauge field $A_1$ is quantized afterwords, using the functional
Schrodinger representation. We derive the anomalous commutators with
nonvanishing Schwinger terms which indicate that our model is
anomalous.

In Section~\ref{sec: const}, we show that the Schwinger term in the
commutator of the Gauss law generators is removed by a redefinition
of these generators and formulate the modified quantum Gauss law
constraint. We prove that this constraint can be also obtained by
using the adiabatic approximation and the notion of quantum holonomy.

In Section~\ref{sec: exoti}, we construct the physical quantum
Hamiltonian consistent with the modified quantum Gauss law constraint,
i.e. invariant under the modified gauge transformations both 
topologically
trivial and non-trivial. We introduce the modified topologically 
non-trivial
gauge transformation operator and define $\theta$--states which are 
its
eigenstates. We consider in detail the case of the SM and demonstrate 
its equivalence to the free field theory of a massive scalar field.
For the generalized CSM, we define the exotic statistics matter field 
and 
reformulate the quantum theory in terms of this field.

In Section~\ref{sec: poinc}, we construct two other Poincare 
generators,
i.e. the momentum and the boost. We act in the same way as before with
the Hamiltonian, namely we define the physical generators as those
which are invariant under both topologically trivial and non-trivial
gauge transformations. We show that the algebra of the constructed
generators is not a Poincare one and that the failure of the Poincare
algebra to close is connected to the nonvanishing vacuum 
Berry curvature.

In Section~\ref{sec: scree}, we study the charge screening. We 
introduce
external charges and calculate $(i)$ the energy of the ground state
of the physical Hamiltonian with the external charges and $(ii)$ the
current density induced by these charges. 

Section~\ref{sec: discu} contains our conclusions and discussion.

\section{Quantization Procedure}
\label{sec: quant}
\subsection{Classical Theory}

The Lagrangian density of the generalized CSM  is
\begin{equation}
{\cal L} = - {\frac{1}{4}} {\rm F}_{\mu \nu} {\rm F}^{\mu \nu} +
\bar{\psi} i {\hbar} {\gamma}^{\mu} {\partial}_{\mu} \psi +
e_{+} \bar{\psi}_{+} {\gamma}^{\mu} {\psi_{+}} A_{\mu} +
e_{-} \bar{\psi}_{-} {\gamma}^{\mu} {\psi_{-}} A_{\mu} ,
\label{eq: odin}
\end{equation}
where ${\rm F}_{\mu \nu}= \partial_{\mu} A_{\nu} - \partial_{\nu} 
A_{\mu}$ ,
$(\mu, \nu) = \overline{0,1}$ , $\gamma^0 = {\sigma}_1$, 
${\gamma}^1 =-i {\sigma}_2$ , ${\gamma}^0 {\gamma}^1 =
{\gamma}^5 = {\sigma}_3$ , ${\sigma}_i$ 
$(i = \overline{1,3})$ are Pauli 
matrices. The field $\psi$ is $2$--component Dirac spinor, 
$\bar{\psi} =
\psi^{\dagger} \gamma^0$ and $\psi_{\pm}=\frac{1}{2} (1 \pm \gamma^5) 
\psi$.

In the temporal gauge $A_0=0$, the Hamiltonian density is
\begin{equation}
{\cal H}  =  {\cal H}_{\rm EM} + {\cal H}_{\rm F},  
\label{eq: dva}
\end{equation}
where
${\cal H}_{\rm EM}  =  \frac{1}{2} {\rm E}^2$, with
${\rm E}$ momentum canonically conjugate to $A_1$, and
\[
{\cal H}_{\rm F} = {\cal H}_{+} + {\cal H}_{-} ,
\]
\[
{\cal H}_{\pm}  \equiv  \psi_{\pm}^{\dagger} d_{\pm} \psi_{\pm} =
\mp \psi_{\pm}^{\dagger}(i{\hbar}{\partial}_{1}+e_{\pm}A_1)\psi_{\pm}.
\]

On the circle boundary conditions for the fields must be specified.
We impose the periodic ones
\begin{eqnarray}
{A_1} (- \frac{\rm L}{2}) & = & {A_1} (\frac{\rm L}{2}) \nonumber \\
{\psi_{\pm}} (- \frac{\rm L}{2}) & = & {\psi_{\pm}} (\frac{\rm L}{2}).
\label{eq: tri}
\end{eqnarray}
We require also that ${\cal H}$ and the classical fermionic currents
$j_{\pm} \equiv {\psi}_{\pm}^{\dagger} {\psi}_{\pm}$ be periodic.

The Lagrangian 
and Hamiltonian densities 
are invariant under local time-independent gauge transformations
\begin{eqnarray*}
A_1 & \rightarrow & A_1 + {\partial}_{1} \lambda,\\
\psi_{\pm} & \rightarrow & \exp\{\frac{i}{\hbar} e_{\pm} \lambda\} 
\psi_{\pm},
\end{eqnarray*}
generated by
\[
{\rm G}  = \partial_{1} {\rm E} + e_{+} j_{+} + e_{-} j_{-},
\]
$\lambda$ being a gauge function, as well as under global gauge
transformations of the right-handed and left-handed Dirac fields 
which are generated by
\[ 
{\rm Q}_{\pm} = e_{\pm} \int_{-{\rm L}/2}^{{\rm L}/2} dx j_{\pm}(x).
\]

Due to the gauge invariance, the Hamiltonian density is not uniquely
determined. On the constrained submanifold ${\rm G} \approx 0$ of the 
full phase space, the Hamiltonian density
\begin{equation}
\tilde{\cal H} = {\cal H} + v_{\rm H} \cdot {\rm G},
\label{eq: shest}
\end{equation}
where $v_{\rm H}$ is an arbitrary Lagrange multiplier which can be 
any function of the field variables and their momenta, reduces 
to the Hamiltonian density ${\cal H}$. In this sense, our theory 
cannot distinguish between ${\cal H}$ and $\tilde{\cal H}$ , and so 
both
Hamiltonian densities are physically equivalent to each other.

For arbitrary $e_{+}, e_{-}$ the gauge transformations do not respect
the boundary conditions ~\ref{eq: tri}. The gauge transformations
compatible with the boundary conditions must be either of the form
\begin{equation}
\lambda (\frac{\rm L}{2})=\lambda (- \frac{\rm L}{2}) + 
{\hbar} {\frac{2\pi}{e_{+}}}n, 
\hspace{5 mm}
{\rm n} \in \cal Z,
\label{eq: sem}
\end{equation}
with $e_{+} \neq 0$ and
\begin{equation}
\frac{e_{-}}{e_{+}} = {\rm N} ,
\hspace{5 mm}
{\rm N} \in \cal Z,
\label{eq: quant}
\end{equation}
or of the form
\[
{\lambda}(\frac{\rm L}{2}) = {\lambda}(-\frac{\rm L}{2}) +
{\hbar} \frac{2\pi}{e_{-}} n,
\hspace{5 mm}
{\rm n} \in \cal Z,
\]
with $e_{-} \neq 0$ and
\begin{equation}
\frac{e_{+}}{e_{-}} = \overline{\rm N} ,
\hspace{5 mm}
\overline{\rm N} \in \cal Z.
\label{eq: cond}
\end{equation}
Eqs. ~\ref{eq: quant} or ~\ref{eq: cond} imply the charge 
quantization
condition for our system. Without loss of generality, we choose the
condition ~\ref{eq: quant}. For ${\rm N}=1$, $e_{-}=e_{+}$ and
we have the standard Schwinger model. For ${\rm N}=0$, we get the
model in which only the right-handed component of the Dirac field
is coupled to the gauge field.

From Eq.~\ref{eq: sem} we see that the gauge transformations under 
consideration 
are divided into topological classes characterized by the integer $n$. 
If
$ \lambda (\frac{\rm L}{2}) = \lambda (- \frac{\rm L}{2})$, then the 
gauge
transformation is topologically trivial and belongs to the $n=0$
class. If $n \neq 0$ it is nontrivial and has winding number $n$.

Given Eq.~\ref{eq: sem}, the nonintegrable phase
\[ 
\Gamma(A)=\exp\{\frac{i}{\hbar} e_{+}\int_{-{\rm L}/2}^{{\rm L}/2}dx 
{A_1}(x,t) \} 
\]
is a unique gauge-invariant quantity that can be constructed from the
gauge field \cite{mant85,raje88,hetr88,sarad88}.
By a topologically trivial transformation we can make
$A_1$ independent of $x$,
\[ {A_1}(x,t) = b(t) ,\]
i.e. obeying the Coulomb gauge ${\partial_1}A_1=0$, then
\[ \Gamma (A) = \exp\{\frac{i}{\hbar} e_{+}{\rm L} b(t)\} .\]
In contrast to $\Gamma (A)$ , the line integral
\[ b(t)={\frac{1}{{\rm L}}}\int_{-{\rm L}/2}^{{\rm L}/2} dx 
{A_1}(x,t) \]
is invariant only under the topologically trivial gauge 
transformations.
The gauge transformations from the $n$th topological class shift $b$
by ${\hbar} \frac{2\pi}{e_{+}{\rm L}}n$. By a non-trivial gauge 
transformation of the
form $g_n=\exp\{i\frac{2\pi}{\rm L}{\hbar}nx\}$, 
we can then bring $b$ into the
interval $[0 , {\hbar}\frac{2\pi}{e_{+}{\rm L}}]$ . 
The configurations $b=0$ and
$b={\hbar}\frac{2\pi}{e_{+}{\rm L}}$ are gauge equivalent, 
since they are connected
by the gauge transformation from the first topological class. The
gauge-field configuration space is therefore a circle with length
${\hbar}\frac{2\pi}{e_{+}{\rm L}}$.

\subsection{Quantization and Anomaly}

The eigenfunctions and the eigenvalues of the first quantized
fermionic Hamiltonians are
\[ 
d_{\pm} \langle x|n;{\pm} \rangle = \pm \varepsilon_{n,{\pm }}
\langle x|n;{\pm } \rangle ,
\]
where
\[ 
\langle x|n;{\pm } \rangle = \frac{1}{\sqrt {\rm L}}
\exp\{\frac{i}{\hbar} e_{\pm} \int_{-{\rm L}/2}^{x} dz{A_1}(z) +
\frac{i}{\hbar} \varepsilon_{n,{\pm}} \cdot x\},  
\]
\[
\varepsilon_{n,{\pm }} = \frac{2\pi}{\rm L} 
(n{\hbar} - \frac{e_{\pm}b{\rm L}}{2\pi}).
\]
We see that the spectrum of the eigenvalues depends on $b$. 
For $\frac{e_{+}b{\rm L}}{2{\pi}{\hbar}}={\rm integer}$, the 
spectrum contains 
the zero energy level. As $b$ increases from $0$ to 
${\hbar}\frac{2\pi}{e_{+}{\rm L}}$, the energies of 
$\varepsilon_{n,+}$ decrease by ${\hbar}\frac{2\pi}{\rm L}$, 
while the energies 
of $(-\varepsilon_{n,-})$ increase by ${\hbar}\frac{2\pi}{\rm L} 
{\rm N}$.
Some of energy levels change sign. However, the spectrum at the 
configurations $b=0$ and $b={\hbar}\frac{2\pi}{e_{+}{\rm L}}$ 
is the same, namely, the integers, as it must be since these 
gauge-field 
configurations are gauge-equivalent. In what follows, we 
will use separately the integer and 
fractional parts of $\frac{e_{+}b{\rm L}}{2{\pi}{\hbar}}$ 
(and $\frac{e_{-}b{\rm L}}{2{\pi}{\hbar}}$ ) , denoting them as 
$[\frac{e_{\pm}b{\rm L}}{2{\pi}{\hbar}}]$ and 
$\{ \frac{e_{\pm}b{\rm L}}{2{\pi}{\hbar}} \}$ 
correspondingly.

Now we introduce the second quantized right-handed and 
left-handed Dirac fields. For the moment, we will assume that 
$d_{\pm}$ 
do not have zero eigenvalues. At time $t=0$, in terms of the 
eigenfunctions of the first quantized fermionic Hamiltonians the 
second 
quantized ($\zeta$--function regulated) fields have the expansion 
\cite{niese86} : 
\[
\psi_{+}^s (x) = \sum_{n \in \cal Z} a_n \langle x|n;{+} \rangle
|\lambda \varepsilon_{n,+}|^{-s/2},
\]
\begin{equation}
\psi_{-}^s (x) = \sum_{n \in \cal Z} b_n \langle x|n;{-} \rangle
|\lambda \varepsilon_{n,-}|^{-s/2}.
\label{eq: vosem}
\end{equation}
Here $\lambda$ is an arbitrary constant with dimension of length
which is necessary to make $\lambda \varepsilon_{n,\pm}$ 
dimensionless,
while $a_n, a_n^{\dagger}$ and $b_n, b_n^{\dagger}$ are 
correspondingly 
right-handed and left-handed fermionic annihilation and creation
operators which fulfil the commutation relations
\[  
[a_n , a_m^{\dagger}]_{+} = [b_n , b_n^{\dagger}]_{+} =\delta_{m,n} .
\]
For $\psi_{\pm }^{s} (x)$, the equal time anticommutators are
\begin{equation}
[\psi_{\pm}^{s}(x) , \psi_{\pm}^{\dagger s}(y)]_{+}=\zeta_{\pm} 
(s,x,y),
\label{eq: devet}
\end{equation}
with all other anticommutators vanishing, where
\[ 
\zeta_{\pm} (s,x,y) \equiv \sum_{n \in \cal Z} \langle x|n;{\pm} 
\rangle
\langle n;{\pm}|y \rangle |\lambda \varepsilon_{n,\pm}|^{-s},
\]
$s$ being large and positive. In the limit, when the regulator
is removed, i.e. $s=0$, $\zeta_{\pm}(s=0,x,y) = \delta(x-y)$ and
Eq.~\ref{eq: devet} takes the standard form.

The vacuum state of the second quantized fermionic Hamiltonian
\[
|{\rm vac};A \rangle = |{\rm vac};A;+ \rangle \otimes
|{\rm vac};A;- \rangle 
\]
is defined such that all negative energy
levels are filled and the others are empty:
\begin{eqnarray}
a_n|{\rm vac};A;+\rangle =0 & {\rm for} & 
n>[\frac{e_{+}b{\rm L}}{2{\pi}{\hbar}}],
\nonumber \\
a_n^{\dagger} |{\rm vac};A;+ \rangle =0 & {\rm for} & n \leq
[\frac{e_{+}b{\rm L}}{2{\pi}{\hbar}}],
\label{eq: deset}
\end{eqnarray}
and
\begin{eqnarray}
b_n|{\rm vac};A;-\rangle =0 & {\rm for} & n \leq
[\frac{e_{-}b{\rm L}}{2{\pi}{\hbar}}], \nonumber \\
b_n^{\dagger} |{\rm vac};A;- \rangle =0 & {\rm for} & n >
[\frac{e_{-}b{\rm L}}{2{\pi}{\hbar}}].
\label{eq: odinodin}
\end{eqnarray}
Excited states are constructed by operating creation operators on the 
Fock vacuum.

In the $\zeta$--function regularization scheme, we define the
action of the functional derivative on first quantized fermionic
kets and bras by
\begin{eqnarray*}
\frac{\delta}{\delta {A_1}(x)} |n;{\pm} \rangle & = & \lim_{s \to 0}
\sum_{m \in \cal Z} |m;{\pm} \rangle \langle m;{\pm}| \frac{\delta}
{\delta {A_1}(x)} |n;{\pm} \rangle \cdot |\lambda \varepsilon_{m,\pm}| 
^{-s/2},\\
\langle n;{\pm }| \frac{\stackrel{\leftarrow}{\delta}}{\delta 
{A_1}(x)}
& = & \lim_{s \rightarrow 0} \sum_{m \in \cal Z} \langle n;{\pm}|
\frac{\stackrel{\leftarrow}{\delta}}{\delta {A_1}(x)} |m;{\pm} \rangle
\langle m;{\pm}| \cdot |\lambda \varepsilon_{m,\pm}|^{-s/2}.
\end{eqnarray*}
From ~\ref{eq: vosem} we get the action of $\frac{\delta}
{\delta {A_1}(x)}$ on the operators $a_n$, $a_n^{\dagger}$
in the form
\begin{eqnarray*}
\frac{\delta}{\delta {A_1}(x)} a_n & = & - \lim_{s \to 0}
\sum_{m \in \cal Z} \langle n;{+}| \frac{\delta}{\delta {A_1}(x)}
|m;{+}\rangle a_m |\lambda \varepsilon_{m,+}|^{-s/2},\\
\frac{\delta}{\delta {A_1}(x)} a_n^{\dagger} & = & \lim_{s \to 0}
\sum_{m \in \cal Z} \langle m;{+}| \frac{\delta}{\delta {A_1}(x)}
|n;{+} \rangle a_m^{\dagger} |\lambda \varepsilon_{m,+}|^{-s/2}.\\
\end{eqnarray*}
The action of $\frac{\delta}{\delta {A_1}(x)}$ on $b_n, b_n^{\dagger}$
can be written analogously.

Next we define the quantum fermionic currents and
fermionic parts of the second-quantized Hamiltonian as
\[
\hat{j}_{\pm}^s(x) = \frac{1}{2} [\psi_{\pm}^{\dagger s}(x),
\psi_{\pm}^{s}(x)]_{-}
\]
and
\[
\hat{\rm H}_{\pm}^s = \int_{-{\rm L}/2}^{{\rm L}/2} dx 
\hat{\cal H}_{\pm}^s(x)= \frac{1}{2} \int_{-{\rm L}/2}^{{\rm L}/2}dx
 (\psi_{\pm}^{\dagger s} d_{\pm} \psi_{\pm}^s
- \psi_{\pm}^s d_{\pm}^{\star} \psi_{\pm}^{\dagger s}).
\]
Substituting ~\ref{eq: vosem} into these expressions, we obtain  
\[
\hat{j}_{\pm}^s(x) =  \sum_{n \in \cal Z} \frac{1}{\rm L}
\exp\{i \frac{2 \pi}{\rm L} nx\} \rho_{\pm}^s(n),
\]
where
\begin{eqnarray*}
\rho_{+}^s(n) & \equiv & \sum_{k \in \cal Z} \frac{1}{2} [a_k^
{\dagger},
a_{k+n}]_{-} \cdot |\lambda \varepsilon_{k,+}|^{-s/2}
|\lambda \varepsilon_{k+n,+}|^{-s/2}, \\
\rho_{-}^s(n) & \equiv & \sum_{k \in \cal Z} \frac{1}{2} [b_k^
{\dagger},
b_{k+n}]_{-} \cdot |\lambda \varepsilon_{k,-}|^{-s/2}
|\lambda \varepsilon_{k+n,-}|^{-s/2}
\end{eqnarray*}
are momentum space charge density (or current) operators, and
\[
\hat{\cal H}_{\pm}^s(x) = \sum_{n \in \cal Z}  \frac{1}{\rm L}
\exp\{i\frac{2\pi}{\rm L}nx\} {\cal H}_{\pm}^s(n), 
\]
\begin{equation}
{\cal H}_{\pm}^s(n) \equiv {\cal H}_{0,\pm}^s(n) \mp 
e_{\pm}b{\rho}_{\pm}^s(n), 
\label{eq: odindva}
\end{equation}
where
\[
{\cal H}_{0,+}^s(n) \equiv {\hbar} \frac{\pi}{\rm L} \sum_{k \in 
\cal Z}
(2k+n) \cdot \frac{1}{2} [a_k^{\dagger}, a_{k+n}]_{-} \cdot
|\lambda \varepsilon_{k,+}|^{-s/2}
|\lambda \varepsilon_{k+n ,+}|^{-s/2} , 
\]
\[
{\cal H}_{0,-}^s(n) \equiv {\hbar} \frac{\pi}{\rm L} \sum_{k \in 
\cal Z}
(2k+n) \cdot \frac{1}{2} [b_{k+n} , b_k^{\dagger}]_{-} \cdot
|\lambda \varepsilon_{k,-}|^{-s/2}
|\lambda \varepsilon_{k+n,-}|^{-s/2} .
\]
The charges corresponding to the currents $\hat{j}_{\pm}^s(x)$ are
\[
\hat{\rm Q}_{\pm}^s = e_{\pm} \int_{-{\rm L}/2}^{{\rm L}/2} dx
\hat{j}_{\pm}^s(x) = e_{\pm} \rho_{\pm}^s(0).
\]
With Eqs.~\ref{eq: deset} and ~\ref{eq: odinodin}, we have for the 
vacuum expectation values:
\begin{eqnarray*}
\langle {\rm vac};A;\pm| \hat{j}_{\pm}(x) |{\rm vac};A;\pm 
\rangle & = &
-\frac{1}{2} \eta_{\pm},\\
\langle {\rm vac},A| \hat{\rm H}_{\rm F} |{\rm vac},A 
\rangle & = &
- \frac{1}{2} (\xi_{+} + \xi_{-}),
\end{eqnarray*}
where
\begin{eqnarray*}
\eta_{\pm} & \equiv & \pm \lim_{s \to 0} \frac{1}{\rm L} 
\sum_{k \in \cal Z} {\rm sign}(\varepsilon_{k,\pm})
|\lambda \varepsilon_{k,\pm}|^{-s},\\
\xi_{\pm} & \equiv & \lim_{s \to 0} \frac{1}{\lambda} 
\sum_{k \in \cal Z} |\lambda \varepsilon_{k,\pm}|^{-s+1}.
\end{eqnarray*}
Taking the sums, we obtain 
\begin{eqnarray*}
{\eta}_{\pm} & = & \pm \frac{2}{\rm L} (\{\frac{e_{\pm}b{\rm L}}
{2{\pi}{\hbar}}\} - \frac{1}{2}), \\
\xi_{\pm} & = & - {\hbar} \frac{2\pi}{\rm L} 
((\{\frac{e_{\pm}b{\rm L}}{2{\pi}{\hbar}}\}
- \frac{1}{2})^2 - \frac{1}{12}).
\end{eqnarray*}

The quantum fermionic currents, charges and Hamiltonians
can be therefore written as
\begin{eqnarray}
\hat{j}_{\pm}(x) & = & :\hat{j}_{\pm}(x):  -  {\frac{1}{2}}
\eta_{\pm},\nonumber \\
\hat{\rm Q}_{\pm} & = & e_{\pm} :\rho_{\pm} (0):  -  
\frac{\rm L}{2} e_{\pm} \eta_{\pm}, \\
\hat{\rm H}_{\pm} & = & \hat{\rm H}_{0,\pm}  \mp  
e_{\pm} b :\rho_{\pm}(0): - \frac{1}{2} \xi_{\pm},   \nonumber
\label{eq: mozol}
\end{eqnarray}
where double dots indicate normal ordering with respect to
$|{\rm vac},A \rangle$ ,
\begin{eqnarray*}
\hat{\rm H}_{0,+} & = & \hbar \frac{2 \pi}{\rm L} \lim_{s \to 0} 
\{ \sum_{k >[\frac{e_{+}b{\rm L}}{2{\pi}{\hbar}}]} k a_k^{\dagger} 
a_k
|\lambda \varepsilon_{k,+}|^{-s}  -  \sum_{k \leq [\frac{e_{+}b
{\rm L}}
{2{\pi}{\hbar}}]} k a_k a_k^{\dagger} |\lambda \varepsilon_
{k,+}|^{-s} \},\\
\hat{\rm H}_{0,-} & = & \hbar \frac{2 \pi}{\rm L} \lim_{s \to 0}
\{ \sum_{k>[\frac{e_{-}b{\rm L}}{2{\pi}{\hbar}}]} k b_{k} b_{k}^
{\dagger}
|\lambda \varepsilon_{k,-}|^{-s} - \sum_{k \leq [\frac{e_{-}b{\rm L}}
{2{\pi}{\hbar}}]} k b_{k}^{\dagger} b_{k}|\lambda \varepsilon_{k,-}|
^{-s} \}
\end{eqnarray*}
and
\begin{eqnarray*}
:\rho_{+} (0): & = & \lim_{s \to 0} \{ \sum_{k >[\frac{e_{+}b{\rm L}}
{2{\pi}{\hbar}}]} a_k^{\dagger} a_k |\lambda \varepsilon_{k,+}|^{-s}-  
\sum_{k \leq [\frac{e_{+}b{\rm L}}{2{\pi}{\hbar}}]} a_k a_k^{\dagger} 
|\lambda \varepsilon_{k,+}|^{-s} \}, \\
:\rho_{-} (0): & = & \lim_{s \to 0} \{ \sum_{k \leq [\frac{e_{-}b
{\rm L}}
{2{\pi}{\hbar}}]} b_{k}^{\dagger}  b_{k} |\lambda \varepsilon_{k,-}|
^{-s} -
\sum_{k>[\frac{e_{-}b{\rm L}}{2{\pi}{\hbar}}]} b_{k} b_{k}^{\dagger}
|\lambda \varepsilon_{k,-}|^{-s} \} . 
\end{eqnarray*}
The operators $:\hat{j}_{\pm}(x):$ and $:\hat{\rm H}_{\pm}:$ are
well defined when acting on finitely excited states which have only a
finite number of excitations relative to the Fock vacuum.

To construct the quantum electromagnetic Hamiltonian, we quantize 
the gauge field using the functional Schrodinger representation.
In this representation, when the vacuum and excited fermionic
Fock states are functionals of $A_1$, the gauge field operators
are represented as $\hat{A}_1(x) \rightarrow A_1(x)$,
$\hat{E}(x) \rightarrow - i{\hbar} \frac{\delta}{\delta A_1(x)}$
and the inner product is evaluated by functional integration.
We first introduce the Fourier expansion for the gauge field
\[
{A_1}(x) = b + \sum_{\stackrel {p \in \cal Z}{p \neq 0}}
e^{i \frac{2 \pi}{\rm L} px} \alpha_p.
\]
Since ${A_1}(x)$ is a real function, $\alpha_p$ satisfies
\[ \alpha_p = \alpha_{-p}^{\star}. \]
The Fourier expansion for the canonical momentum conjugate to 
${A_1}(x)$ is then
\[
\hat{\rm E}(x) = \frac{1}{\rm L} \hat{\pi}_b - \frac{i}{\rm L} \hbar
\sum_{\stackrel {p \in \cal Z} {p \neq 0}} 
e^{-i \frac{2\pi}{\rm L}px} \frac{d}{d{\alpha_p}} ,
\]
where $\hat{\pi}_b \equiv -i \hbar \frac{d}{db}$ .
The electromagnetic part of the Hamiltonian density is
\[
\hat{\cal H}_{\rm EM}(x) = \sum_{p \in \cal Z} \frac{1}{\rm L}
\exp\{i\frac{2\pi}{\rm L}px\} \cdot {\cal H}_{\rm EM}(p),
\]
where
\begin{equation}
{\cal H}_{\rm EM}(p) \equiv - \frac{1}{\rm L} {\hbar}^2
\frac{d}{d{\alpha}_{-p}} \frac{d}{db} - \frac{1}{2\rm L} {\hbar}^2
\sum_{\stackrel{q \in \cal Z}{q \neq (0;p)}}
\frac{d}{d{\alpha}_{-p+q}} \frac{d}{d{\alpha}_{-q}}
\hspace{1 cm} (p \neq 0) ,
\label{eq: odincet}
\end{equation}
so the corresponding quantum Hamiltonian becomes
\[
\hat{\rm H}_{\rm EM} = {\cal H}_{\rm EM}(p=0) = \frac{1}{2\rm L} 
\hat{\pi}_b^2 - \frac{1}{\rm L} {\hbar}^2 \sum_{q >0} 
\frac{d}{d{\alpha_q}} \frac{d}{d{\alpha_{-q}}}.
\]
The total quantum Hamiltonian is
\[
\hat{\rm H} = \hat{\rm H}_{0,+} + \hat{\rm H}_{0,-} + 
\hat{\rm H}_{\rm EM}
-e_{+}b :\rho_{+}(0): + e_{-}b :\rho_{-}(0): - 
\frac{1}{2} (\xi_{+} + \xi_{-}).
\]

If we multiply two operators that are finite linear combinations of
the fermionic creation and annihilation operators, the $\zeta$--
function
regulated operator product agrees with the naive product. However, if
the operators involve infinite summations their naive product is not
generally well defined. We then define the operator product by
mutiplying the regulated operators with $s$ large and positive and
analytically continue the result to $s=0$. In this way we obtain the
following relations  
\begin{equation}
[\rho_{\pm}(m) , \rho_{\pm}(n)]_{-} = \pm m \delta_{m,-n}, 
\label{eq: odinpet}
\end{equation}
\[
[{\cal H}_{0,\pm}(n), {\cal H}_{0,\pm}(m)]_{-}=
\pm {\hbar} \frac{2\pi}{\rm L} (n-m) {\cal H}_{0,\pm}(n+m),
\]
\[
[\hat{\rm H}_{0,\pm} ,\rho_{\pm}(m)]_{-} = 
\mp \hbar \frac{2 \pi}{\rm L} m \rho_{\pm} (m),
\]              
and
\[
\frac{d}{db}  \rho_{\pm}(m) =  0,
\]
\begin{eqnarray}
\frac{d}{d{\alpha_{\pm p}}} \rho_{+}(m) & = & 
-\frac{e_{+}{\rm L}}{2{\pi}{\hbar}}
\delta_{p, \pm m} ,  \nonumber \\     
\frac{d}{d{\alpha_{\pm p}}} \rho_{-}(m) & = & 
\frac{e_{-}{\rm L}}{2 {\pi}{\hbar}}
\delta_{p, \pm m}, \hspace{1 cm}  (p >0).
\label{eq: odinshest}
\end{eqnarray}
 
The quantum Gauss operator is
\[
\hat{\rm G} = \hat{\rm G}_0 + \frac{2 \pi}{{\rm L}^2}
\sum_{p>0} \{ \hat{\rm G}_{+}(p) e^{i\frac{2 \pi}{\rm L} px} -
\hat{\rm G}_{-}(p) e^{-i\frac{2\pi}{\rm L} px} \},
\]
where
\begin{eqnarray*}
\hat{\rm G}_0 & \equiv & \frac{1}{\rm L} e_{+} \rho_{\rm N}(0) , \\
\hat{\rm G}_{\pm}(p) & \equiv & \hbar p\frac{d}{d{\alpha}_{\mp p}} \pm
\frac{e_{+}\rm L}{2\pi} \rho_{\rm N}(\pm p) 
\end{eqnarray*}
and $\rho_{\rm N}=\rho_{+} + {\rm N} \rho_{-}$ is momentum space
total charge density operator.

Using ~\ref{eq: odinpet} and ~\ref{eq: odinshest}, we easily get
that $\rho_{+}(\pm p)$ ( and $\rho_{-}(\pm p)$ ) are 
gauge invariant. For example, for $\rho_{+}(\pm p)$ we have:
\[
[\hat{\rm G}_{+}(p), \rho_{+}(\pm q) ]_{-} = 0,       
\]
\[
[\hat{\rm G}_{-}(p), \rho_{+}(\pm q) ]_{-} = 0, 
\]
$(p>0, q>0)$. The operators $\hat{\rm G}_{\pm}(p)$ don't commute 
with themselves,
\[
[\hat{\rm G}_{+}(p) , \hat{\rm G}_{-}(q)]_{-} =
(1-{\rm N}^2) \frac{e_{+}^2{\rm L}^2}{4{\pi}^2} p\delta_{p,q}
\]
as well as with the Hamiltonian
\[
[\hat{\rm H} , \hat{\rm G}_{\pm}(p)]_{-} =
\pm (1-{\rm N}^2) {\hbar} \frac{e_{+}^2{\rm L}}{4{\pi}^2} 
\frac{d}{d{\alpha_{\mp p}}},  \hspace{1 cm}  (p>0).  
\]
The last two commutators reflect an anomalous behaviour of the 
generalized CSM. The appearance of the Schwinger term in the first
commutator changes the nature of the Gauss law constraints: instead
of being first class constraints, they turn into second class ones.
The Schwinger term in the second commutator means that the total
quantum Hamiltonian is not invariant under the topologically trivial
gauge transformations generated by $\hat{\rm G}_{\pm}(p)$.

For ${\rm N}=1$, i.e. for the standard SM, both commutators vanish.
Another case of vanishing Schwinger terms is axial electrodynamics
where ${\rm N}=-1$ and the fermionic fields ${\psi_{\pm}}$ are of
opposite charge.

\section{Quantum Constraints}
\label{sec: const}

\subsection{Quantum Symmetry}

In non-anomalous gauge theories, Gauss law is considered to be valid
for physical states only. This identifies physical states as those
which are gauge-invariant. The problem with the anomalous behaviour
of the generalized CSM, in terms of states in Hilbert space, is 
apparent: 
owing to the Schwinger terms we cannot require
that states be annihilated by the Gauss law generators $\hat{\rm G}_
{\pm}(p)$.

Let us represent the action of the topologically trivial gauge
transformations by the operators
\begin{equation}
{\rm U}_{0}(\tau) = \exp\{\frac{i}{\hbar} \hat{\rm G}_{0}
{\tau}_0 + \frac{i}{\hbar} \sum_{p>0}
(\hat{\rm G}_{+} \tau_{+} + \hat{\rm G}_{-} \tau_{-}) \}
\label{eq: odinsem}
\end{equation}
with $\tau_0$ , ${\tau}_{\pm}(p)$ smooth, then 
\begin{eqnarray*}
{\rm U}_0^{-1}(\tau) {\alpha}_{\pm p} {\rm U}_0(\tau) & = &
{\alpha}_{\pm}  - i p {\tau}_{\mp}(p),\\
{\rm U}_0^{-1}(\tau) \frac{d}{d {\alpha}_{\pm p}} {\rm U}_0(\tau) &=&
\frac{d}{d {\alpha}_{\pm p}} \mp 
\frac{i}{{\hbar}^2} (1-{\rm N}^2) 
(\frac{e_{+}\rm L}{2\pi})^2 {\tau}_{\pm}(p),
\hspace{5 mm} (p>0).
\end{eqnarray*}
The composition law for the operators ${\rm U}_{0}$ is
\[
{\rm U}_0({\tau}^{(1)}) {\rm U}_0({\tau}^{(2)}) =
\exp\{ 2\pi i {\omega}_{2}({\tau}^{(1)}, {\tau}^{(2)})\}
{\rm U}_{0}( {\tau}^{(1)} + {\tau}^{(2)} ) ,
\]
where
\[
{\omega}_{2}({\tau}^{(1)} , {\tau}^{(2)}) \equiv
- \frac{i}{4\pi} (1-{\rm N}^2) (\frac{e_{+}{\rm L}}{2{\pi}{\hbar}})^2 
\sum_{p>0} p 
({\tau }_{-}^{(1)} {\tau }_{+}^{(2)} - {\tau}_{+}^{(1)} {\tau}_{-}^
{(2)})
\]
is a 2-cocycle of the gauge group algebra. Thus for ${\rm N} 
\neq \pm 1$ 
we are dealing with a projective representation.

The 2-cocycle ${\omega}_{2}( {\tau}^{(1)} , {\tau}^{(2)} )$
is trivial, since it can be removed   by
a simple redefinition of ${\rm U}_{0}(\tau)$. Indeed, the modified
operators
\begin{equation}
\tilde{\rm U}_0(\tau) = \exp\{i 2\pi {\alpha}_{1}(\gamma ; \tau)\}
\cdot {\rm U}_{0}(\tau),
\label{eq: odinvosem}
\end{equation}
where
\[
{\alpha}_{1}(\gamma, \tau) \equiv - \frac{1}{4 \pi} (1-{\rm N}^2)
(\frac{e_{+}{\rm L}}{2{\pi}{\hbar}})^2
\sum_{p>0}( {\alpha}_{-p}{\tau}_{-} - {\alpha}_{p} {\tau}_{+} )
\]
is a 1-cocycle, satisfy the ordinary composition law
\[
\tilde{\rm U}_0({\tau}^{(1)}) \tilde{\rm U}_0({\tau}^{(2)}) =
\tilde{\rm U}_0({\tau}^{(1)} + {\tau}^{(2)}),
\]
i.e. the action of the topologically trivial gauge transformations
represented by ~\ref{eq: odinvosem} is unitary.

The modified Gauss law generators corresponding to ~\ref{eq: 
odinvosem}
are 
\begin{equation}
\hat{\tilde{\rm G}}_{\pm}(p) = \hat{\rm G}_{\pm}(p)  \pm 
\frac{1}{\hbar} (1-{\rm N}^2)\frac{{e_{+}^2}{\rm L}^2}{8{\pi}^2}
{\alpha}_{\pm p}.
\label{eq: odindevet}
\end{equation}
The generators $\hat{\tilde{\rm G}}_{\pm}(p)$ commute:
\[
[\hat{\tilde{\rm G}}_{+}(p) , \hat{\tilde{\rm G}}_{-}(q)]_{-}=0.
\]
This means that Gauss law can be maintained at the quantum level 
for ${\rm N} \neq \pm 1$, too. We define physical  states as those 
which 
are annihilated by $\hat{\tilde{\rm G}}_{\pm}(p)$   {\cite{sarad91}}:
\begin{equation}
\hat{\tilde{\rm G}}_{\pm}(p) |{\rm phys}; A \rangle = 0.
\label{eq: dvanol}
\end{equation}
The zero component ${\hat{\rm G}}_0$ is a sum of quantum generators 
of the global gauge transformations of the right-handed and 
left-handed fermionic fields, so the other quantum constraints are
\begin{equation}
:\rho_{\pm}(0): |{\rm phys};A \rangle = 0.
\label{eq: dvaodin}
\end{equation}
It follows from ~\ref{eq: dvanol} that the physical states
$|{\rm phys};A \rangle$ respond to a gauge transformation from
the zero topological class with a phase:
\begin{equation}
{\rm U}_0(\tau) |{\rm phys};A \rangle =
\exp\{-i 2\pi {\alpha}_1({\gamma};\tau) \} |{\rm phys};A \rangle.
\label{eq: zvezda}
\end{equation}
Only for models without anomaly, i.e. for ${\rm N}= \pm 1$, this
equation translates into the statement that physical states are
gauge invariant.

Equation ~\ref{eq: zvezda} expresses in an exact form the nature
of anomaly in the CSM. At the quantum level the gauge
invariance is not broken , but realized projectively.
The $1$-cocycle ${\alpha}_1$ occuring in the projective
representation contributes to the commutator of the Gauss
law generators by a Schwinger term and produces therefore
the anomaly.

\subsection{Adiabatic Approach}

Let us show now that we can come to the quantum constraints
~\ref{eq: dvanol} and ~\ref{eq: dvaodin} in a different way, using
the adiabatic approximation  \cite{schiff68,berry84}.
In the adiabatic approach, the dynamical variables are divided
into two sets, one which we call fast variables and the other
which we call slow variables. In our case, we treat the fermions
as fast variables and the gauge fields as slow variables.

Let ${\cal A}^1$ be a manifold of all static gauge field
configurations ${A_1}(x)$. On ${\cal A}^1$  a time-dependent
gauge field ${A_1}(x,t)$ corresponds to a path and a periodic gauge
field to a closed loop.

We consider the fermionic part of the second-quantized Hamiltonian
$:\hat{\rm H}_{\rm F}:$ which depends on $t$  through the background
gauge field $A_1$ and so changes very slowly with time. We consider
next the periodic gauge field ${A_1}(x,t) (0 \leq t <T)$ . After a
time $T$ the periodic field ${A_1}(x,t)$ returns to its original
value: ${A_1}(x,0) = {A_1}(x,T)$, so that $:\hat{\rm H}_{\rm F}:(0)=
:\hat{\rm H}_{\rm F}:(T)$ .

At each instant $t$ we define eigenstates for $:\hat{\rm H}_{\rm F}:
(t)$ by
\[
:\hat{\rm H}_{\rm F}:(t) |{\rm F}, A(t) \rangle =
{\varepsilon}_{\rm F}(t) |{\rm F}, A(t) \rangle.
\]
The state $|{\rm F}=0, A(t) \rangle \equiv |{\rm vac}, A(t) \rangle$
is a ground state of $:\hat{\rm H}_{\rm F}:(t)$ ,
\[
:\hat{\rm H}_{\rm F}:(t) |{\rm vac}, A(t) \rangle =0. 
\]
The Fock states $|{\rm F}, A(t) \rangle $ depend on $t$ only through
their implicit dependence on $A_1$. They are assumed to be 
periodic in time, $|{\rm F}, A(T) \rangle = |{\rm F}, A(0) \rangle$,
orthonormalized,
\[
\langle {\rm F^{\prime}}, A(t)|{\rm F}, A(t) \rangle =
\delta_{{\rm F},{\rm F^{\prime}}},    
\]
and nondegenerate.

The time evolution of the wave function  of our system (fermions
in a background gauge field) is clearly governed by the Schrodinger
equation:
\[ 
i \hbar \frac{\partial \psi(t)}{\partial t} =
:\hat{\rm H}_{\rm F}:(t) \psi(t) .  
\]
For each $t$, this wave function can be expanded in terms of the
"instantaneous" eigenstates $|{\rm F}, A(t) \rangle$ .

Let us choose ${\psi}_{\rm F}(0)=|{\rm F}, A(0) \rangle$, i.e.
the system is initially described by the eigenstate
$|{\rm F},A(0) \rangle$ . According to the adiabatic approximation,
if at $t=0$ our system starts in an stationary state $|{\rm F},A(0)
\rangle $ of $:\hat{\rm H}_{\rm F}:(0)$, then it will remain,
at any other instant of time $t$, in the corresponding eigenstate
$|{\rm F}, A(t) \rangle$ of the instantaneous Hamiltonian
$:\hat{\rm H}_{\rm F}:(t)$. In other words, in the adiabatic
approximation transitions to other eigenstates are neglected.

At time  $t=T$
our system will be described by the state
\[
{\psi}_{\rm F}(T) = \exp\{i {\gamma}_{\rm F}^{\rm dyn} +
i {\gamma}_{\rm F}^{\rm Berry} \}\cdot {\psi}_{\rm F}(0),  
\]
where
\[ {\gamma}_{\rm F}^{\rm dyn} \equiv - \frac{1}{\hbar}
\int_{0}^{T} dt \cdot {\varepsilon}_{\rm F}(t) ,    \]
while
\begin{equation}
{\gamma}_{\rm F}^{\rm Berry} \equiv \int_{0}^{T} dt \int_{-{\rm L}/2}^
{{\rm L}/2} dx \dot{A_1}(x,t) \langle {\rm F},A(t)|i \frac{\delta}
{\delta A_1(x,t)}|{\rm F},A(t) \rangle
\label{eq: dvadva}
\end{equation}
is Berry's phase  \cite{berry84}.

If we define the $U(1)$ connection
\begin{equation}
{\cal A}_{\rm F}(x,t) \equiv \langle {\rm F},A(t)|i \frac{\delta}
{\delta A_1(x,t)}|{\rm F},A(t) \rangle,
\label{eq: dvatri}
\end{equation}
then
\[
{\gamma}_{\rm F}^{\rm Berry} = \int_{0}^{T} dt \int_{-{\rm L}/2}^
{{\rm L}/2} dx \dot{A}_1(x,t) {\cal A}_{\rm F}(x,t).   
\]
We see that upon parallel transport around a closed loop on
${\cal A}^1$ the Fock state $|{\rm F},A(t) \rangle$ acquires an 
additional phase which is integrated exponential of ${\cal A}_{\rm F}
(x,t)$. Whereas the dynamical phase ${\gamma}_{\rm F}^{\rm dyn}$
provides information about the duration of the evolution, the
Berry's phase reflects the nontrivial holonomy of the Fock states
on ${\cal A}^1$.

However, a direct computation of the diagonal matrix elements of
$\frac{\delta}{\delta A_1(x,t)}$ in ~\ref{eq: dvadva} requires a
globally single-valued basis for the eigenstates $|{\rm F},A(t) 
\rangle$
which is not available \cite{jpha}. 
For that reason, to calculate ${\gamma}_{\rm F}^{\rm Berry}$ it
is more convenient to compute first the $U(1)$ curvature tensor
\begin{equation}
{\cal F}_{\rm F}(x,y,t) \equiv \frac{\delta}{\delta A_1(x,t)}
{\cal A}_{\rm F}(y,t) - \frac{\delta}{\delta A_1(y,t)}
{\cal A}_{\rm F}(x,t)
\label{eq: dvacet}
\end{equation}
and then deduce ${\cal A}_{\rm F}$.

The vacuum curvature tensor 
is evaluated as \cite{jpha}
\begin{equation}
{\cal F}_{{\rm F}=0} = 
(1-{\rm N}^2) \frac{e_{+}^2}{2{\pi}^2{\hbar}^2} \sum_{n>0} \frac{1}{n}
\sin(\frac{2\pi}{\rm L} n(x-y)) = 
(1-{\rm N}^2) \frac{e_{+}^2}{2\pi{\hbar}^2} ( \frac{1}{2} 
\epsilon(x-y)
- \frac{1}{\rm L} (x-y) ).
\label{eq: dvasem}
\end{equation}
The corresponding $U(1)$ connection is easily deduced as
\[
{\cal A}_{{\rm F}=0}(x,t) = -\frac{1}{2} \int_{-{\rm L}/2}
^{{\rm L}/2} dy {\cal F}_{{\rm F}=0}(x,y,t) A_1(y,t).
\]
The Berry phase becomes
\[
{\gamma}_{{\rm F}=0}^{\rm Berry} = - \frac{1}{2} \int_{0}^{T} dt
\int_{-{\rm L}/2}^{{\rm L}/2} dx \int_{-{\rm L}/2}^{{\rm L}/2} dy
\dot{A_1}(x,t) {\cal F}_{{\rm F}=0}(x,y,t) A_1(y,t).  
\]

In terms of the Fourier components, the connection ${\cal A}_
{{\rm F}=0}$ is rewritten as
\begin{eqnarray*}
\langle {\rm vac},A(t)|  \frac{d}{db(t)}  | {\rm vac},A(t) \rangle &
= & 0,\\
\langle {\rm vac},A(t)|  \frac{d}{d{\alpha}_{\pm p}(t)}  |
{\rm vac},A(t) \rangle & \equiv & {\cal A}_{\pm}(p,t) =
\pm (1-{\rm N}^2) \frac{e_{+}^2{\rm L}^2}{8{\pi}^2{\hbar}^2} 
\frac{1}{p} {\alpha}_{\mp p},
\end{eqnarray*}
so the nonvanishing curvature is
\[
{\cal F}_{+ -}(p) \equiv \frac{d}{d{\alpha}_{-p}} {\cal A}_{+} -
\frac{d}{d{\alpha}_{p}} {\cal A}_{-} = (1-{\rm N}^2) 
\frac{e_{+}^2{\rm L}^2}{4{\pi}^2{\hbar}^2} \frac{1}{p} .    
\]
A parallel transportation of the vacuum $|{\rm vac},A(t) \rangle$ 
around a closed loop in $({\alpha}_{p}, {\alpha}_{-p})$ -- 
space $(p>0)$ yields back the same vacuum state multiplied by the 
phase
\[
{\gamma}_{{\rm F}=0}^{\rm Berry} = (1-{\rm N}^2)
\frac{e_{+}^2{\rm L}^2}{4{\pi}^2{\hbar}^2} 
\int_{0}^{T} dt \sum_{p>0}\frac{1}{p} 
i{\alpha}_{p} \dot{\alpha}_{-p}.    
\]
This phase is associated with the projective representation
of the gauge group. For ${\rm N} = \pm 1$, when the representation
is unitary, the curvature ${\cal F}_{+-}$ and the Berry phase vanish.

As mentioned in the beginning of this Section, the projective
representation is trivial  and the 2-cocycle in the composition
law of the gauge transformation operators can be removed by a 
redefinition  of these operators. Analogously, if we redefine 
the momentum operators as
\begin{equation}
\frac{d}{d{\alpha}_{\pm p}} \longrightarrow 
\frac{\tilde{d}}{d{\alpha}_{\pm p}} \equiv \frac{d}{d{\alpha}_{\pm p}} 
\mp (1-{\rm N}^2) \frac{e_{+}^2{\rm L}^2}{8{\pi}^2{\hbar}^2} 
\frac{1}{p} {\alpha}_{\mp p},
\label{eq: dvavosem}
\end{equation}
then the corresponding connection and curvature vanish:
\begin{eqnarray*}
\tilde{{\cal A}}_{\pm} & \equiv & \langle {\rm vac},A(t)|
\frac{\tilde{d}}{d{\alpha}_{\pm p}} |{\rm vac},A(t) \rangle =0,\\
\tilde{\cal F}_{+ -} & = & \frac{\tilde{d}}{d{\alpha}_{-p}}
\tilde{\cal A}_{+} - \frac{\tilde{d}}{d{\alpha}_{p}}
\tilde{\cal A}_{-} =0.
\end{eqnarray*}
However, the nonvanishing curvature ${\cal F}_{+-}(p)$ shows itself
in the algebra of the modified momentum operators  which are 
noncommuting:
\[
[\frac{\tilde{d}}{d{\alpha}_{p}} , \frac{\tilde{d}}
{d{\alpha}_{-q}} ]_{-} ={\cal F}_{+-}(p) {\delta}_{p,q}.
\]
Following ~\ref{eq: dvavosem}, we modify the Gauss law generators
as
\[
\hat{\rm G}_{\pm}(p) \longrightarrow \hat{\tilde{\rm G}}_{\pm}(p) =
\hbar p \frac{\tilde{d}}{d{\alpha}_{\mp p}} \pm 
\frac{e_{+}\rm L}{2\pi} \rho_{\rm N}(\pm p) 
\]
that coincides with ~\ref{eq: odindevet}. The modified Gauss law
generators have vanishing vacuum  expectation values,
\[
\langle {\rm vac},A(t)| \hat{\tilde{\rm G}}_{\pm}(p,t) |
{\rm vac},A(t) \rangle =0.  
\]
This justifies the definition ~\ref{eq: dvanol}.

For the zero component $\hat{\rm G}_0$, the vacuum expectation
value
\[
\langle {\rm vac},A(t)| \hat{\rm G}_0 | {\rm vac},A(t) \rangle
= - \frac{1}{2} ( e_{+} {\eta}_{+} + e_{-} {\eta}_{-})    
\]
can be also made equal to zero by the redefinition
\[
\hat{\rm G}_0 \longrightarrow \hat{\tilde{\rm G}}_0 =
\hat{\rm G}_0 + \frac{1}{2} (e_{+} {\eta}_{+} +
e_{-} {\eta}_{-}) = \frac{1}{\rm L}
e_{+} :\rho_{\rm N}(0):      
\]
that leads to ~\ref{eq: dvaodin}.

Thus, both quantum constraints ~\ref{eq: dvanol} and
~\ref{eq: dvaodin} can be realized in the framework of the
adiabatic approximation.

\section{Physical Quantum CSM}
\label{sec: exoti}
\subsection{Construction of Physical Hamiltonian}

$1$. From the point of view of Dirac quantization, there are many
physically equivalent classical theories of a system with
first-class constraints. The origin of such an
ambiguity lies in a gauge freedom. For the classical CSM, the
gauge freedom is characterized by an arbitrary $v_{\rm H}(x)$
in ~\ref{eq: shest}. If we use the Fourier expansion for
$v_{\rm H}(x)$, then the general form of the classical Hamiltonian   
is rewritten as
\begin{equation}
\tilde{\rm H} = {\rm H} + 
\sum_{p>0} ( v_{\rm H,+} {\rm G}_{+} + v_{\rm H,-} {\rm G}_{-}) .
\label{eq: dvadevet}
\end{equation}
Any Hamiltonian $\tilde{\rm H}$ with fixed nonzero
$(v_{\rm H,-} , v_{\rm H,+})$ gives rise to the same weak equations
of motion as those deduced from ${\rm H}$, although the
strong form of these equations may be quite different. The physics is
however described by the weak equations. Different $(v_{\rm H,-},
v_{\rm H,+})$ lead to different mathematical descriptions of the
same physical situation.

To construct the quantum theory of any system with first-class
constraints, we usually quantize one of the corresponding classical
theories. All the possible quantum theories constructed in this way
are believed to be equivalent to each other.

In the case, when gauge degrees of freedom are anomalous, the
situation is different: the physical equivalence of quantum
Hamiltonians is lost. For the CSM, the quantum Hamiltonian
$\hat{\tilde{\rm H}}$ does not reduce to $\hat{\rm H}$
on the physical states:
\[
\hat{\tilde{\rm H}} |{\rm phys};A \rangle \neq
\hat{\rm H} |{\rm phys};A \rangle .     
\]
The quantum theory consistently describing the dynamics of the CSM
should be definitely compatible with ~\ref{eq: dvanol}. The
corresponding quantum Hamiltonian is then defined by the conditions
\begin{equation}
[\hat{\tilde{\rm H}} , \hat{\tilde{\rm G}}_{\pm}(p)]_{-} =0
\hspace{1 cm} (p>0)
\label{eq: trinol}
\end{equation}
which specify that $\hat{\tilde{\rm H}}$ must be invariant
under the modified topologically trivial gauge transformations
generated by $\hat{\tilde{\rm G}}_{\pm}(p)$.

We have in ~\ref{eq: trinol} a system  of non-homogeneous
equations in the Lagrange multipliers $\hat{v}_{\rm H,\pm}$
which become operators at the quantum level. The solution of
these equations is 
\[
\hat{v}_{\rm H,\pm}(p) = \frac{\hbar}{\rm L} \frac{1}{p^2}
\{ p \frac{d}{d{\alpha}_{\pm p}} \mp 
(1-{\rm N}^2) (\frac{e_{+}{\rm L}}{4{\pi}{\hbar}})^2 {\alpha}_{\mp p} 
\}.     
\]
Substituting this expression for $\hat{v}_{\rm H,\pm}(p)$ 
into the quantum counterpart of ~\ref{eq: dvadevet}, on the physical 
states $|{\rm phys};A \rangle$ we obtain
\[
\frac{1}{2} \sum_{p>0} \{ [\hat{v}_{\rm H,+}(p), \hat{\rm G}_{+}
(p)]_{+} 
+ [\hat{v}_{\rm H,-}(p) , \hat{\rm G}_{-}(p)]_{+} \} = 
\frac{1}{{\rm L}^2} {\hbar}^2 \sum_{p>0} (\frac{d}{d{\alpha}_{p}}
\frac{d}{d{\alpha}_{-p}} - \frac{1}{2} [\frac{\tilde{d}}
{d{\alpha}_{p}},
\frac{\tilde{d}}{d{\alpha}_{-p}}]_{+} ),
\]
i.e. the last term in the right-hand side of ~\ref{eq: dvadevet}
contributes only to the electromagnetic part of the Hamiltonian,
changing $\frac{d}{d{\alpha}_{\pm}}$ by $\frac{\tilde{d}}
{d{\alpha}_{\pm}}$:
\[
\hat{\rm H}_{\rm EM} \rightarrow \hat{\tilde{\rm H}}_{\rm EM}=
\tilde{\cal H}_{\rm EM}(0) \equiv
\frac{1}{2\rm L} \hat{\pi}_{b}^{2} -
\frac{1}{2\rm L} {\hbar}^2 \sum_{p>0}[\frac{\tilde{d}}{d{\alpha}_{p}},
\frac{\tilde{d}}{d{\alpha}_{-p}}]_{+} .     
\]
In terms of the momentum space charge density operators, the gauge
invariant electromagnetic Hamiltonian becomes
\[
\hat{\tilde{\rm H}}_{\rm EM} = \frac{1}{2\rm L} \hat{\pi}^2_{b} +
{\rm V}({\rho}_{\rm N};{\rho}_{\rm N}),
\]
where
\[
{\rm V}({\rho}_{\rm N};{\rho}_{\rm N}) \equiv
\frac{e_{+}^2 {\rm L}}{8{\pi}^2}
\sum_{\stackrel{p \in \cal Z}{p \neq 0}} \frac{1}{p^2}
{\rho}_{\rm N}(-p) {\rho}_{\rm N}(p)
\]
is the energy of the Coulomb current-current interaction.

In order to make the dependence on $\rm N$ for the Hamiltonian
more obvious, let us represent ${\rho}_{\rm N}$ as
\[
{\rho}_{\rm N} = \frac{1}{2} (1+{\rm N}) {\rho} +
\frac{1}{2} (1-{\rm N}) {\sigma},
\]
where
\begin{eqnarray*}
{\rho} \equiv {\rho}_1 & = & {\rho}_{+} + {\rho}_{-},\\
{\sigma} \equiv {\rho}_{-1} & = & {\rho}_{+} - {\rho}_{-},
\end{eqnarray*}
and
\[
[{\rho}(p),{\rho}(q)]_{-} = [{\sigma}(p),{\sigma}(q)]_{-}=0,
\]
\[
[{\sigma}(p),{\rho}(q)]_{-} = 2p {\delta}_{p,-q}.
\]
Then the Coulomb interaction energy takes the form
\begin{equation}
{\rm V}({\rho}_{\rm N};{\rho}_{\rm N}) = \frac{1}{4} (1+{\rm N})^2
{\rm V}({\rho};{\rho}) + \frac{1}{4} (1-{\rm N})^2
{\rm V}({\sigma};{\sigma}) + \frac{1}{2} (1-{\rm N}^2)
{\rm V}({\rho};{\sigma}).
\label{eq: triodin}
\end{equation}
For ${\rm N}=1$, ${\rho}(p)$ and ${\sigma}(p)$ are respectively
momentum space electric and axial charge density operators, the
electromagnetic Hamiltonian depending only on ${\rho}$:
\[
\hat{\tilde{\rm H}}_{\rm EM} = \frac{1}{2\rm L} \hat{\pi}^2_{b} +
{\rm V}({\rho};{\rho}).
\]
For ${\rm N}=-1$, the momentum space electric charge density operator
is ${\sigma}(p)$ and
\[
\hat{\tilde{\rm H}}_{\rm EM} = \frac{1}{2\rm L} \hat{\pi}^2_b +
{\rm V}({\sigma};{\sigma}).
\]
For ${\rm N} \neq \pm 1$, i.e. for models with anomaly, the last term
in ~\ref{eq: triodin} does not vanish and is of principal importance.
This term means that ${\rho}$ and ${\sigma}$ are not decoupled  as
before for the cases without anomaly and that the electromagnetic
Hamiltonian involves the non-commuting charge density operators.

$2$. The topologically nontrivial gauge transformations change the 
integer
part of $\frac{e_{+}b{\rm L}}{2{\pi}{\hbar}}$ :
\begin{eqnarray*}
[\frac{e_{+}b{\rm L}}{2{\pi}{\hbar}}] & \rightarrow & 
[\frac{e_{+}b{\rm L}}{2{\pi}{\hbar}}]+n,\\
\hat{\psi}_{+} & \rightarrow & \exp\{i
\frac{2{\pi}n}{\rm L} x\}
\hat{\psi}_{+},
\end{eqnarray*}    
and
\begin{eqnarray*}
[\frac{e_{-}b{\rm L}}{2{\pi}{\hbar}}] & \rightarrow & 
[\frac{e_{-}b{\rm L}}{2{\pi}{\hbar}}]+
{\rm N} \cdot n, \\
\hat{\psi}_{-} & \rightarrow & \exp\{i {\rm N}
\frac{2{\pi}n}{\rm L}x\}
\hat{\psi}_{-}.
\end{eqnarray*}
The action of the topologically nontrivial gauge transformations on
the states can be represented by the operators
\begin{equation}
{\rm U}_n = \exp\{-\frac{i}{\hbar} n \cdot \hat{\rm T}_{b}\}\cdot 
{\rm U}_0
\label{eq: tridva}
\end{equation}
where
\[
\hat{\rm T}_{b} \equiv \hat{\pi}_{[\frac{e_{+}b{\rm L}}{2{\pi}
{\hbar}}]} -
\frac{2\pi}{\rm L} \int_{-{\rm L}/2}^{{\rm L}/2} dx x \cdot
(\hat{j}_{+}(x) + {\rm N} \hat{j}_{-}(x)) \equiv
-i \hbar \frac{d}{d[\frac{e_{+}b{\rm L}}{2{\pi}{\hbar}}]} + i{\hbar} 
\sum_{\stackrel{n \in \cal Z}{n \neq 0}} \frac{(-1)^n}{n} 
\rho_{\rm N}(n)    
\]
and ${\rm U}_0$ is given by ~\ref{eq: odinsem}.

To identify the gauge transformation as belonging to the $n$th
topological class we use the index $n$ in ~\ref{eq: tridva}. The
case $n=0$ corresponds to the topologically trivial gauge
transformations.

The topologically nontrivial gauge transformation operators satisfy
the same composition law as the topologically trivial ones.
The modified operators are
\[
\tilde{\rm U}_n = \exp\{- \frac{i}{\hbar} n \cdot \hat{\rm T}_b\}
\cdot \tilde{\rm U}_0.      
\]
On the physical states
\[
\tilde{\rm U}_n |{\rm phys};A \rangle = 
(\exp\{-\frac{i}{\hbar}\hat{\rm T}_{b} \})^n |{\rm phys};A \rangle.          
\]

Among all states $|{\rm phys};A \rangle$ one may identify the
eigenstates of the operators of the physical variables. The action
of the topologically nontrivial gauge transformations on such
states may, generally speaking, change only the phase of these
states by a C--number, since with any gauge transformations both
topologically trivial and nontrivial, the operators of the physical
variables and the observables cannot be changed. Using
$|{\rm phys}; \theta \rangle$ to designate these physical states,
we have
\[
\exp\{\mp \frac{i}{\hbar} \hat{\rm T}_b\}  |{\rm phys}; \theta \rangle 
= e^{\pm i\theta}  |{\rm phys}; \theta \rangle.
\]
The states $|{\rm phys};\theta \rangle$ 
are easily constructed in the form
\[
|{\rm phys};\theta \rangle = \sum_{n \in \cal Z} e^{-in\theta}
(\exp\{-\frac{i}{\hbar} \hat{\rm T}_b\})^n |{\rm phys};A \rangle
\]
(so called $\theta$--states \cite{jack76,callan76}),
where $|{\rm phys};A \rangle$ is an arbitrary physical state from
~\ref{eq: dvanol}.

In one dimension the parameter $\theta$ is related to a constant
background electric field . To show this, let us  introduce states 
which
are invariant even against the topologically nontrivial gauge
transformations. Recalling that $[\frac{e_{+}b{\rm L}}{2{\pi}
{\hbar}}]$ 
is shifted
by $n$ under a gauge transformation from the $n$th topological class,
we obtain 
such states by the following transition 
\begin{equation}
|{\rm phys}; \theta \rangle  \rightarrow
|{\rm phys}\rangle \equiv \exp\{i[\frac{e_{+}b{\rm L}}{2{\pi}
{\hbar}}] 
\theta\} |{\rm phys};\theta \rangle.
\label{eq: tritri}
\end{equation}
The new states $|{\rm phys} \rangle$ continue to be annihilated by
$\hat{\tilde{\rm G}}_{\pm}(p)$, and are also invariant under
the topologically nontrivial gauge transformations. 

The electromagnetic part of the
Hamiltonian transforms as 
\[
\hat{\rm H}_{\rm EM} \rightarrow
\exp\{i[\frac{e_{+}b{\rm L}}{2{\pi}{\hbar}}]\theta\} \hat{\rm H}_
{\rm EM}
\exp\{-i[\frac{e_{+}b{\rm L}}{2{\pi}{\hbar}}]\theta\}
\]
\[
=\frac{1}{2\rm L}
(\hat{\pi}_b - {\rm L} {\cal E}_{\theta})^2 -
\frac{1}{2\rm L} {\hbar}^2 \sum_{p>0} 
[\frac{\tilde{d}}{d{\alpha}_{p}},\frac{\tilde{d}}{d{\alpha}_{-p}}
]_{+},      
\]
i.e. in the new Hamiltonian the momentum $\hat{\pi}_b$ is supplemented 
by 
the electric 
field strength ${\cal E}_{\theta} \equiv \frac{e_{+}}{2\pi} \theta$.

$3.$The Fourier components of the fermionic currents are transformed
under the topologically non-trivial gauge transformations
as follows
\begin{eqnarray*}
\rho_{+}(\pm p) & \rightarrow & \rho_{+}(\pm p) 
- (-1)^p \cdot n, \\
\rho_{-}(\pm p) & \rightarrow & \rho_{-}(\pm p) 
+ (-1)^p \cdot
{\rm N} \cdot n, 
\hspace{5 mm} (p>0),
\end{eqnarray*}
being invariant under the topologically trivial ones.

The quantum Hamiltonian invariant under the topologically trivial
gauge transformations is still not uniquely determined. 
We can add to it any linear
combination of the operators  $\rho_{+}(\pm p)$ and
$\rho_{-}(\pm p)$ :
\begin{equation}
\hat{\tilde{\rm H}} \rightarrow \hat{\tilde{\rm H}} +
\beta_{0} + \sum_{\stackrel{p \in \cal Z}{p \neq 0}} 
( \beta_{+} \cdot \rho_{+}(p) + \beta_{-} \cdot \rho_{-}(p)) 
\label{eq: tripet}
\end{equation}
where $\beta_{0}$, $\beta_{\pm}$  are arbitrary functions.
The conditions ~\ref{eq: trinol} does not clearly fix these functions.

The Hamiltonian of the consistent quantum theory of the generalized
CSM should be invariant under the topologically nontrivial gauge
transformations as well. So next to ~\ref{eq: trinol} is the
following condition
\begin{equation}
[\hat{\tilde{\rm H}} , \hat{\rm T}_b]_{-} =0.
\label{eq: trishest}
\end{equation}
The condition ~\ref{eq: trishest} can be  then rewritten as a
system of linear equations in $(\beta_{0},\beta_{\pm})$. 
We can easily find a solution of these equations, which gives
us $(\beta_{0},\beta_{\pm})$ as functions of 
$[\frac{e_{+}b{\rm L}}{2{\pi}{\hbar}}]$ . 
The most general solution must involve constants depending on
$\{\frac{e_{\pm}b{\rm L}}{2{\pi}{\hbar}}\}$. However, these constants
are irrelevant for our consideration and we neglect them.

Finding $(\beta_{0},\beta_{\pm})$ from ~\ref{eq: trishest} and
substituting them into the expression ~\ref{eq: tripet}, on the
physical states we obtain
\[
\hat{\tilde{\rm H}} |{\rm phys};A \rangle =
\hat{\rm H}_{\rm phys} |{\rm phys};A \rangle
\]
where
\[
\hat{\rm H}_{\rm phys} = \hat{\rm H}_{\rm F}^{\rm phys} +
\hat{\rm H}_{\rm EM}^{\rm phys},
\]
\[ 
\hat{\rm H}_{\rm F}^{\rm phys}  = 
\hat{\rm H}_{0,+} + \hat{\rm H}_{0,-}
-\frac{1}{2} (\xi_{+} + \xi_{-})
- \frac{\pi}{\rm L} {\hbar} (1+{\rm N}^2)
([\frac{e_{+}b{\rm L}}{2{\pi}{\hbar}}])^2  
\]
\begin{equation}
+ \frac{2\pi}{\rm L}{\hbar} 
[\frac{e_{+}b{\rm L}}{2{\pi}{\hbar}}] 
\sum_{\stackrel{p \in \cal Z}{p \neq 0}}
(-1)^p {\rho}_{-{\rm N}}(p),
\label{eq: hamil}
\end{equation}
\[
\hat{\rm H}_{\rm EM}^{\rm phys} =  \frac{1}{2\rm L} \hat{\pi}^2_b +
{\rm V}({\rho}_{\rm N};{\rho}_{\rm N}) +
\frac{e_{+}^2{\rm L}}{4{\pi}^2} (1-{\rm N}^2)
[\frac{e_{+}b{\rm L}}{2{\pi}{\hbar}}] 
\sum_{\stackrel{p \in \cal Z}{p \neq 0}} 
\frac{(-1)^p}{p^2} \rho_{\rm N}(p) 
\]
\begin{equation}
+  \frac{e_{+}^2{\rm L}}{24} (1-{\rm N}^2)^2 
([\frac{e_{+}b{\rm L}}{2{\pi}{\hbar}}])^2 .
\label{eq: trisem}
\end{equation}
The free fermionic Hamiltonians $\hat{\rm H}_{0,\pm}$ can be
expressed in terms of ${\rho}_{\pm}(p)$, by making use of the
bosonization procedure. Their bosonized version is
\[
\hat{\rm H}_{0,\pm}^s = \frac{2\pi}{\rm L} \hbar
\sum_{p>0} |{\lambda}{\varepsilon}_{p,\pm}|^{-s}
{\rho}_{\pm}^s(-p) {\rho}_{\pm}^s(p).
\]
Eqs. ~\ref{eq: hamil} and ~\ref{eq: trisem}  
give us a physical Hamiltonian invariant
under both topologically trivial and nontrivial gauge transformations,
$\hat{\rm H}_{\rm F}^{\rm phys}$ and $\hat{\rm H}_{\rm EM}^{\rm phys}$
being invariant separately.
The last two terms in ~\ref{eq: hamil} make invariant the free 
fermionic
part of the Hamiltonian, while the ones in ~\ref{eq: trisem} the
electromagnetic part. 

For ${\rm N} = \pm 1$, the last two terms in ~\ref{eq: trisem} vanish.
These terms are therefore caused by the anomaly and   represent new
types of interaction which are absent in the nonanomalous models.

The new interactions admit the following interpretation. 
Let us combine the last term in ~\ref{eq: trisem} with the kinetic
part of the electromagnetic Hamiltonian, then
\[
\frac{1}{2\rm L} \hat{\pi}^2_b +  
\frac{e_{+}^2\rm L}{24}
(1-{\rm N}^2)^2 ([\frac{e_{+}b{\rm L}}{2{\pi}{\hbar}}])^2 =
\frac{1}{2{\rm L}^2} \int_{-{\rm L}/2}^{{\rm L}/2} dx
(\hat{\pi}_b - {\rm L} {\cal E}(x))^2 ,
\]
i.e. the momentum $\hat{\pi}_b$ is supplemented by the
linearly rising electric field strength
\[
{\cal E}(x) \equiv - \frac{{e_{+}}}{\rm L} x
(1-{\rm N}^2) [\frac{e_{+}b{\rm L}}{2{\pi}{\hbar}}].       
\]
As in four-dimensional models of a relativistic particle moving
in an external field, we may define a generalized momentum operator
in the form
\[
\hat{\tilde{\pi}}_b(x) \equiv \hat{\pi}_b - {\rm L}
{\cal E}(x).
\]
The commutation relations for $\hat{\tilde{\pi}}_b$ are
\[
[\hat{\tilde{\pi}}_b(x), \hat{\tilde{\pi}}_b(y)]_{-} =
i (1-{\rm N}^2) \frac{e_{+}^2{\rm L}}{2{\pi}} (x-y).
\]
We see that due to the new interactions the physical degrees of 
freedom behave themselves as moving in a background linearly
rising electric field . This is an effective field not related
directly to the original fields of our model. It may be 
considered as
produced by a charge uniformly distributed on the circle with 
density
\[
{\rho}_{\rm bgrd} = - \frac{1}{{\rm L}} (1-{\rm N}^2)
[\frac{e_{+}b{\rm L}}{2{\pi}{\hbar}}].
\]

This situation is similar to that in $(2+1)-$ or $(3+1)-$dimensions.
As known, in the nonAbelian models governed by Lagrangians with
topological terms (the Pontryagin density in $(3+1)$-dimensions
or the Chern--Simons term in $(2+1)$-dimensions) the nonAbelian
gauge field is moving in a background ${\rm U}(1)$ functional gauge
potential expressed in terms of the nonAbelian gauge field components
\cite{jack83}. The peculiarity of the situation in our case is that
there is no magnetic field related to the gauge field in $(1+1)$-
dimensions, so the background field is electric.

If we make again the transition to the physical states invariant
under both the topologically trivial and nontrivial gauge
transformations, then the density of the kinetic part of the
physical electromagnetic Hamiltonian becomes:
\[
\frac{1}{2{\rm L}^2} \hat{\tilde{\pi}}_{b}^2 \rightarrow 
\frac{1}{2{\rm L}^2}
(\hat{\pi}_b - {\rm L} ({\cal E}_{\theta} +
{\cal E}(x)))^2.   
\]
While the constant background electric field is general in
one-dimensional gauge models defined on the circle, the linearly
rising one is specific to models with left-right asymmetric
matter content \cite{sarad92} .

The next-to-last term in ~\ref{eq: trisem} means that the fermionic
physical degrees of freedom and $b$ are not decoupled in the physical
Hamiltonian. This term represents the Coulomb type background-matter
interaction: 
\[
\frac{e_{+}^2{\rm L}}{4{\pi}^2} (1-{\rm N}^2)
[\frac{e_{+}b{\rm L}}{2{\pi}{\hbar}}] 
\sum_{\stackrel{p \in \cal Z}{p \neq 0}}
\frac{(-1)^p}{p^2} {\rho}_{\rm N}(p) =-
\frac{e_{+}^2{\rm L}^2}{4{\pi}^2}
\sum_{\stackrel{p \in \cal Z}{p \neq 0}} \frac{(-1)^p}{p^2}
{\rho}_{\rm bgrd} \cdot {\rho}_{\rm N}(p).
\]
It is just the background linearly rising electric field that
couples $b$ to the fermionic physical degrees of freedom in
the Coulomb interaction.

As a consequence, the eigenstates of the physical Hamiltonian
are not a direct product of the purely fermionic Fock states
and wave functionals of $b$. This is a common feature of gauge
theories with anomaly. That the Hilbert space in such theories
is not a tensor product of the Hilbert space for a gauge field
and the fixed Hilbert space for fermions was shown in
\cite{nels85}, \cite{fadd86}.

The background charge interpretation is related to the definition
of the Fock vacuum. The definition given in Eqs.~\ref{eq: deset}-
~\ref{eq: odinodin} depends on $[\frac{e_{+}b{\rm L}}{2\pi\hbar}]$
and remains unchanged only locally on ${\cal A}^1$, in regions
where $[\frac{e_{+}b{\rm L}}{2\pi\hbar}]$ is fixed. The values
of the gauge field in regions of different $[\frac{e_{+}b{\rm L}}
{2\pi\hbar}]$ are connected by the topologically nontrivial gauge
transformations. If $[\frac{e_{+}b{\rm L}}{2\pi\hbar}]$ changes,
then there is a nontrivial spectral flow, i.e. some of energy
levels of the first quantized fermionic Hamiltonians cross zero
and change sign. This means that the definition of the Fock
vacuum changes.

The charge operators $\hat{\rm Q}_{\pm}$ also change. Let
$:\hat{\rm Q}_{\pm}^{(0)}:$ be charge operators defined in the
region where $[\frac{e_{+}b{\rm L}}{2\pi\hbar}]=0$ and 
normal ordered with respect to the corresponding Fock vacuum.
Then in regions with nonzero $[\frac{e_{+}b{\rm L}}{2\pi\hbar}]$
the charge operators become $:\hat{\rm Q}_{\pm}^{(0)}: \mp e_{\pm}
[\frac{e_{\pm}b{\rm L}}{2\pi\hbar}]$ . For models without anomaly,
the additional terms in the positive and negative chirality
charges are opposite in sign, so the total charge is
$:\hat{\rm Q}_{+}^{(0)}: + :\hat{\rm Q}_{-}^{(0)}:$ in all
regions of different $[\frac{e_{+}b{\rm L}}{2\pi\hbar}]$, i.e.
defined globally on ${\cal A}^1$. For models with anomaly,       
the additional terms do not cancel each other and the total
charge operator up to terms depending on $\{\frac{e_{\pm}b
{\rm L}}{2\pi\hbar}\}$ becomes $:\hat{\rm Q}_{+}^{(0)}: +
:\hat{\rm Q}_{-}^{(0)}: + e_{+}{\rm L}{\rho}_{\rm bgrd}$.
The background charge is therefore that part of the total
charge which depends on $[\frac{e_{+}b{\rm L}}{2\pi\hbar}]$
and changes in the transition between regions of different
$[\frac{e_{+}b{\rm L}}{2\pi\hbar}]$.

\subsection{Exotization}
We can
formally decouple the matter and gauge field degrees of freedom by
introducing the exotic statistics matter fields  ( the so-called
exotization procedure \cite{sarad94}). Let us define the composite 
fields
\begin{equation}
\tilde{\psi}_{\pm}(x) = \exp\{\mp i \frac{\pi}{\rm L} x \pm
i \frac{2\pi}{e_{\pm}{\rm L}} 
(\hat{\pi}_{b} \mp e_{\pm} x [\frac{e_{\pm}b {\rm L}}{2{\pi}{\hbar}}]
)\}
\cdot {\psi}_{\pm}(x) .
\label{eq: cetodin}
\end{equation}
The fields $\tilde{\psi}_{\pm}(x)$ are invariant under the 
topologically nontrivial gauge transformations (we put
${\rm U}_0=1$ )
\[
\exp\{\frac{i}{\hbar} n \hat{\rm T}_b \} \tilde{\psi}_{\pm}
\exp\{- \frac{i}{\hbar} n \hat{\rm T}_b \} =
\tilde{\psi}_{\pm}
\]
and have the commutation relations
\begin{eqnarray}
\tilde{\psi}_{\pm}^\dagger (x) \tilde{\psi}_{\pm}(y) & + &
e^{\mp i{\rm F}(x,y)}  \tilde{\psi}_{\pm}(y)  
\tilde{\psi}_{\pm}^{\dagger}(x)  =  \delta (x-y), \nonumber \\  
\vspace{5 mm}
\tilde{\psi}_{\pm}(x)  \tilde{\psi}_{\pm}(y) & + & 
e^{\pm i{\rm F}(x,y)}  \tilde{\psi}_{\pm}(y)  \tilde{\psi}_{\pm}
(x)  = 0,
\label{eq: cetdva}
\end{eqnarray}
where ${\rm F}(x,y) \equiv \frac{2\pi}{\rm L}(x-y)$ .
The commutation relations ~\ref{eq: cetdva} are indicative of an
exotic statistics of $\tilde{\psi}_{\pm}(x)$. These fields are neither
fermionic nor bosonic. Only for $x=y$ Eqs.~\ref{eq: cetdva} become
anti-commutators: $\tilde{\psi}_{\pm}(x)$ ( and $\tilde{\psi}_{\pm}
^{\dagger}(x)$ ) anticommute with themselves , i.e. behave as  
fermionic fields.

Using ~\ref{eq: cetodin} and 
the expansions ~\ref{eq: vosem}, we obtain the Fourier
expansions for the exotic fields $\tilde{\psi}_{\pm}(x)$ :
\[
\tilde{\psi}_{+}^{s}(x) = \sum_{n \in \cal Z} \tilde{a}_n
\langle x|n;{+} \rangle |\lambda \varepsilon_{n,+}|^{-s/2}, 
\]
\[
\tilde{\psi}_{-}^{s}(x) = \sum_{n \in \cal Z} \tilde{b}_n
\langle x|n;{-} \rangle |\lambda \varepsilon_{n,-}|^{-s/2},
\]
where
\[
\tilde{a}_n  \equiv  \exp\{i \frac{2\pi}{e_{+}{\rm L}}
\hat{\pi}_b\} a_{n+[\frac{e_{+}b{\rm L}}{2{\pi}{\hbar}}]} ,
\]
\[
\tilde{b}_n \equiv \exp\{-i \frac{2\pi}{e_{-}{\rm L}}
\hat{\pi}_b \} b_{n+[\frac{e_{-}b{\rm L}}{2{\pi}{\hbar}}]} .
\]
The exotic creation and annihilation operators $\tilde{a}_n^{\dagger},
\tilde{a}_n$ and $\tilde{b}_n^{\dagger} , \tilde{b}_n$ fulfil the 
following commutation relations algebra:
\begin{eqnarray*}
\tilde{a}_n^{\dagger} \tilde{a}_m   +  \tilde{a}_{m-1} 
\tilde{a}_{n-1}^{\dagger} & = & \delta_{mn}, \\  
\tilde{a}_n  \tilde{a}_m  + \tilde{a}_{m+1} \tilde{a}_{n-1} & = & 0,
\end{eqnarray*}
and
\begin{eqnarray*}
\tilde{b}_n^{\dagger} \tilde{b}_m + \tilde{b}_{m+1}
\tilde{b}_{n+1}^{\dagger} & = & \delta_{mn}, \\
\tilde{b}_n \tilde{b}_m + \tilde{b}_{m-1} \tilde{b}_{n+1} & = & 0.
\end{eqnarray*}
We next introduce the new Fock vacuum $\overline{|{\rm vac};A 
\rangle}=
\overline{|{\rm vac};A;+ \rangle} \bigotimes 
\overline{|{\rm vac};A;- \rangle}$
defined as
\begin{eqnarray*}
a_n \overline{|{\rm vac};A;+ \rangle}=0 
& {\rm for} & n>0, \\
a_n^{\dagger} \overline{|{\rm vac};A;+ \rangle}=0 
& {\rm for} & n \leq 0,
\end{eqnarray*}
and
\begin{eqnarray*}
b_n \overline{|{\rm vac};A;- \rangle}=0
& {\rm for} & n \leq 0, \\ 
b_n^{\dagger} \overline{|{\rm vac};A;- \rangle}=0
& {\rm for} & n>0,
\end{eqnarray*}
denoting the normal ordering with respect to $\overline{|{\rm vac};A
\rangle}$ by $\vdots \hspace{2 mm} \vdots$ .

If we compare the old and the new definitions of the Fock vacuum,
then we see a shift of the level that separates the filled leves
and the empty ones. The new Fock vacuum is defined such that
for all values of $[\frac{e_{+}b{\rm L}}{2\pi\hbar}]$ only the
levels with energy lower than (or equal to) the energy of the
level $n=0$ are filled and the others are empty, i.e. the new
definition does not depend on $[\frac{e_{+}b{\rm L}}{2\pi\hbar}]$
and remains unchanged as the gauge configuration changes.

The exotic matter current operators are  
\[
\hat{\tilde{j}}_{\pm}^{s}(x)  =  \sum_{n \in \cal Z}
\frac{1}{\rm L} \exp\{i \frac{2\pi}{\rm L} nx\} \cdot 
\tilde{\rho}_{\pm}^s(n),
\]
\[
\tilde{\rho}_{+}^s(n)  =  \sum_{k \in \cal Z} 
\tilde{a}_{k}^{\dagger}  \tilde{a}_{k+n} \cdot
|\lambda \varepsilon_{k,+}|^{-s/2} |\lambda \varepsilon_{k+n,+}|
^{-s/2},  
\]
\[
\tilde{\rho}_{-}^s(n) = \sum_{k \in \cal Z}
\tilde{b}_{k}^{\dagger} \tilde{b}_{k-n} \cdot
|\lambda \varepsilon_{k,-}|^{-s/2} |\lambda \varepsilon_{k-n,-}|
^{-s/2}.
\]
The new operators $\tilde{\rho}_{\pm}(n)$ and the old ones 
$\rho_{\pm}(n)$ are connected in the following way:
\[
\vdots \tilde{\rho}_{\pm}(n) \vdots = :\rho_{\pm}(n): 
\pm \delta_{n,0} [\frac{e_{\pm}b{\rm L}}{2{\pi}{\hbar}}]. 
\]
The exotic matter charges are
\[
\vdots \hat{\tilde{\rm Q}}_{\pm} \vdots = :\hat{\rm Q}_{\pm}: 
\pm e_{\pm} [\frac{e_{\pm}b{\rm L}}{2{\pi}{\hbar}}].       
\]
On the physical states ~\ref{eq: dvaodin} the exotic charges
become
\begin{equation}
\vdots \hat{\tilde{\rm Q}}_{\pm} \vdots |{\rm phys};A\rangle 
=\pm e_{\pm} [\frac{e_{\pm}b{\rm L}}{2{\pi}{\hbar}}] |{\rm phys};A 
\rangle.
\label{eq: cettri}
\end{equation}

With ~\ref{eq: cettri}, we decouple the matter and gauge-field 
degrees 
of freedom in the physical Hamiltonian ~\ref{eq: trisem}. We obtain
\[
\hat{\rm H}_{\rm phys} = \frac{1}{2\rm L} \hat{\pi}_b^2 -
\frac{1}{2}  (\xi_{+} + \xi_{-}) +
\hbar \frac{2\pi}{\rm L} \sum_{p>0}
(\rho_{\rm tot,+}(-p) \rho_{\rm tot,+}(p) 
+ \rho_{\rm tot,-}(p) \rho_{\rm tot,-}(-p) )
\]
\[
+ {\rm V}({\rho}_{\rm N}^{\rm tot}; {\rho}_{\rm N}^{\rm tot}),
\] 
where we have defined the operators
\[
\rho_{\rm N}^{\rm tot} \equiv \rho_{{\rm tot},+} + 
{\rm N} \cdot {\rho}_{{\rm tot},-} ,
\]
\[
{\rho}_{{\rm tot},\pm} \equiv \tilde{\rho}_{\pm} + (-1)^p
\frac{1}{e_{\pm}} \vdots \hat{\tilde{\rm Q}}_{\pm} \vdots.
\]
These operators are invariant under both topologically trivial and 
nontrivial gauge transformations.

To diagonalize the exotic matter part of the physical Hamiltonian,
we perform the Bogoliubov transformation over the operators 
$\rho_{\rm tot,+}(\pm p)$ and $\rho_{\rm tot,-}(\pm p)$ ,
$(p>0)$ : 
\begin{eqnarray}
{\rho}_{\rm tot,+}(\pm p) & \rightarrow & \overline{\rho}_{\rm tot,+}
(\pm p)
= \cosh{t_p} \cdot {\rho}_{\rm tot,+}(\pm p) +
\sinh{t_p} \cdot {\rho}_{\rm tot,-}(\pm p), \nonumber \\ 
{\rho}_{\rm tot,-}(\pm p) & \rightarrow & \overline{\rho}_{\rm tot,-}
(\pm p)
= \sinh{t_p} \cdot {\rho}_{\rm tot,+}(\pm p) +
\cosh{t_p} \cdot {\rho}_{\rm tot,-}(\pm p),
\label{eq: bogo}
\end{eqnarray}
where 
\begin{eqnarray*}
\cosh{2t_p} & = & \frac{1}{{\rm E}_p} (\frac{2{\pi}p}{\rm L}\hbar +
\frac{e_{+}^2{\rm L}}{8{\pi}^2 p}  (1+ {\rm N}^2) ), \\
\sinh{2t_p} & = & \frac{1}{{\rm E}_p} 
\frac{e_{+}^2{\rm L}}{4{\pi}^2 p} {\rm N},
\end{eqnarray*}
and
\[
{\rm E}_p = \sqrt{{\rm E}_p^2({\rm N})
+ (\frac{e_{+}^2{\rm L}}{8{\pi}^2})^2
(1-{\rm N}^2)^2 \frac{1}{p^2}},
\]
\[
{\rm E}_p^2({\rm N}) \equiv (\frac{2{\pi}p}{\rm L})^2 {\hbar}^2+ 
\frac{e_{+}^2}{2\pi} \hbar (1+ {\rm N}^2).
\]
The Bogoliubov transformed operators
$\overline{\rho}_{\rm tot,+}(\pm p) , \overline{\rho}_{\rm tot,-}(\pm 
p)$ satisfy the same commutation relations as the nontransformed ones:
\[
[\overline{\rho}_{\rm tot,+}(m) , \overline{\rho}_{\rm tot,+}(n)]_{-}=
[\overline{\rho}_{\rm tot,-}(n) , \overline{\rho}_{\rm tot,-}(m)]_{-}=
m {\delta}_{m,-n}.
\]
The generator of the Bogoliubov transformation ~\ref{eq: bogo} is
\[
{\rm B}_p \equiv \exp\{\frac{1}{p} t_p (\overline{\rho}_{\rm tot,-}(p)
\overline{\rho}_{\rm tot,+}(-p) - \overline{\rho}_{\rm tot,+}(p)
\overline{\rho}_{\rm tot,-}(-p)) \}.
\]
The diagonalized form of the total physical Hamiltonian is
\begin{equation}
\hat{\rm H}_{\rm phys} = \frac{1}{2\rm L} \hat{\pi}_b^2 -
\frac{1}{2} (\xi_{+} + \xi_{-}) + \sum_{p>0}
\frac{1}{p} {\rm E}_p ( \overline{\rho}_{\rm tot,+}(-p)
\overline{\rho}_{\rm tot,+}(p) + 
\overline{\rho}_{\rm tot,-}(p) \overline{\rho}_{\rm tot,-}(-p) ).
\label{eq: diogo}
\end{equation}
The physical Hamiltonian obtained is expressed in terms of the
exotic matter and global gauge-field degrees of freedom. The exotic
fields are composites of the fermionic matter and background
electric fields. 

For the ${\rm N}=1$ model, $e_{-}=e_{+} \equiv e$ and the linearly 
rising
background electric field vanishes. The spectrum of the physical
Hamiltonian becomes relativistic
\[
{\rm E}_p = {\rm E}_p({\rm N}=1)=
{\hbar} \sqrt{(\frac{2{\pi}p}{\rm L})^2 + {\rm M}^2},
\]
where ${\rm M}^2 \equiv \frac{e^2}{\pi}\frac{1}{\hbar}$.

If we introduce the creation and annihilation operators for $b$,
\begin{eqnarray*}
{\rm C}^{\dagger} & \equiv & \frac{1}{\sqrt{2{\rm ML}}}
(-{\hbar} \frac{d}{db} + 2\sqrt{\pi \hbar}(\{\frac{eb{\rm L}}
{2{\pi}{\hbar}} \} - \frac{1}{2} ) ),\\
{\rm C} & \equiv & \frac{1}{\sqrt{2{\rm ML}}} 
({\hbar} \frac{d}{db} +2\sqrt{\pi \hbar} (\{\frac{eb{\rm L}}
{2{\pi}{\hbar}} \} - \frac{1}{2} ) ),\\
\end{eqnarray*}
\[
[{\rm C} , {\rm C}^{\dagger}] =1,
\]
then the global gauge-field part of the physical Hamiltonian becomes
\[
\frac{1}{2\rm L} \hat{\pi}_b^2 - \frac{1}{2} (\xi_{+} + \xi_{-})
= {\rm M} ({\rm C}^{\dagger} {\rm C} + \frac{\hbar}{2} ).
\]
The wave function of its lowest energy eigenstate is
\[
f_0(b) = (\frac{{\rm M}{\rm L}}{{\pi}{\hbar}})^{1/4} 
\exp\{ -({\frac{2\pi}{e{\rm L}}})^2 \frac{{\rm ML}{\hbar}}{2} 
(\{\frac{eb{\rm L}}{2{\pi}{\hbar}}\} -\frac{1}{2})^2\}.
\]
The total physical Hamiltonian takes the form
\begin{equation}
\hat{\rm H}_{\rm phys} = {\rm M}{\rm C}^{\dagger} {\rm C} +
\sum_{p>0} \frac{1}{p} {\rm E}_p
(\overline{\rho}_{\rm tot,+}(-p) \overline{\rho}_{\rm tot,+}(p) +
\overline{\rho}_{\rm tot,-}(p) \overline{\rho}_{\rm tot,-}(-p)).
\label{eq: total}
\end{equation}
This is just the Hamiltonian of a massive scalar boson with mass
${\rm M}$.

The ${\rm N} =-1$ model can be considered analogously. We get the
same physical Hamiltonian ~\ref{eq: total}. Thus, for both cases
${\rm N} = \pm 1$, the quantum generalized CSM is equivalent to the
free field theory of a massive scalar field.

For the ${\rm N} \neq \pm 1$ models, the spectrum of the physical
Hamiltonian is nonrelativistic and does not correspond to a massive
boson. So the quantum theory of the models with anomaly is not
equivalent to the theory of a free massive scalar field.
 
Since the matter and gauge-field degrees of freedom are decoupled
in the physical Hamiltonian ~\ref{eq: diogo}, its eigenstates can be 
represented as a direct product of the exotic matter Fock states and 
wave
functionals of $b$. In particular, the ground state of the physical
Hamiltonian is defined as
\[
(\frac{1}{2\rm L} \hat{\pi}_b^2  - \frac{1}{2} 
(\xi_{+} + \xi_{-})) |{\rm ground} \rangle =0,
\]
\[
\overline{\rho}_{\rm tot,+}(n) |{\rm ground} \rangle =
\overline{\rho}_{\rm tot,-}(-n) |{\rm ground} \rangle =0,
\hspace{1 cm} 
n>0.
\]
For the ${\rm N}=1$ model, the ground state is
\[
|{\rm ground} \rangle = f_0(b) \cdot (\prod_{n>0} {\rm U}_n^{\dagger})
\overline{|{\rm vac};{\rm A} \rangle}\\
\]
\[
= f_0(b) \cdot \exp\{ - \sum_{n>0} \frac{1}{n} 
( \overline{\rho}_{\rm tot,-}(n) \overline{\rho}_{\rm tot,+}(-n) -
\overline{\rho}_{\rm tot,+}(n) \overline{\rho}_{\rm tot,-}(-n)) \} 
\cdot
\overline{|{\rm vac};{\rm A} \rangle}.\\
\]
All the excited states are constructed by acting the Bogoliubov 
transformed operators \\
$\overline{\rho}_{\rm tot,+}(-n),
\overline{\rho}_{\rm tot,-}(n), (n>0)$ and the global gauge-field
degree of freedom creation operator ${\rm C}^{\dagger}$ on \\
the ground state.

Thus, the quantum generalized CSM
can be formulated in two equivalent ways.
In the first way, the matter fields are fermionic and coupled 
nontrivially
to the global gauge-field degree of freedom. In the second way, 
the matter
and gauge-field degrees of freedom are decoupled in the   physical
Hamiltonian, but the matter fields acquire an exotic statistics. 

\section{Poincare Algebra}
\label{sec: poinc}
$1$. The classical momentum and boost generators are given by
\[
{\rm P}=\int_{-{\rm L}/2}^{{\rm L}/2} dx
(-i\hbar \psi_{+}^{\dagger} \partial_1 \psi_{+} 
- i\hbar \psi_{-}^{\dagger} \partial_1 \psi_{-} - {\rm E} \partial_1 
A), 
\]
\[
{\rm K} =  \int_{-{\rm L}/2}^{{\rm L}/2} dx \cdot x {\cal H}(x) .
\]
After a straightforward calculation we obtain 
\[
\{ {\rm H} , {\rm P} \} = 0 , 
\]
\[
\{ {\rm P} , {\rm K} \} = - {\rm H} , 
\hspace{5 mm} 
\{ {\rm H} , {\rm K} \} = - {\rm P}, 
\]
i.e. at the classical level, these generators obey the Poincare
algebra.

At the quantum level, the momentum and boost generators become
\[
\hat{\rm P}= \hat{\rm P}_{+} + \hat{\rm P}_{-} -
\int_{-{\rm L}/2}^{{\rm L}/2} dx \hat{\rm E} {\partial_1}A_1 ,
\]
\begin{eqnarray*}
\hat{\rm P}_{\pm} & \equiv & \frac{1}{2} \hbar \int_{-{\rm L}/2}
^{{\rm L}/2} dx ( {\psi}_{\pm}^{\dagger} (-i\partial_1) {\psi}_{\pm}
- {\psi}_{\pm} (i\partial_1) {\psi}_{\pm}^{\dagger}), \\
\hat{\rm K} & = & \int_{-{\rm L}/2}^{{\rm L}/2} dx \cdot x
(\hat{\cal H}_{+}(x) + \hat{\cal H}_{-}(x) + \hat{\cal H}_{\rm EM}
(x)).
\end{eqnarray*}
Using the Fourier expansions for the fermionic and gauge fields,
we rewrite  the quantum generators as
\[
\hat{\rm P} = \hat{\rm P}_{+} - \hat{\rm P}_{-} -
\frac{e_{+}^2 \rm L}{2 \pi} (1-{\rm N}^2)
\sum_{p > 0} {\alpha}_{-p}{\alpha}_{p} ,
\]
\[
\hat{\rm P}_{\pm} = \pm \hat{\rm H}_{0,\pm} \mp \frac{1}{2} 
{\xi}_{\pm} - \frac{1}{2} e_{\pm} {\eta}_{\pm} b{\rm L},
\]
\[
\hat{\rm K} = -i \frac{\rm L}{2\pi} \sum_{\stackrel{p \in \cal Z}
{p \neq 0}} \frac{(-1)^p}{p} ({\cal H}_{+}(p) + {\cal H}_{-}(p) +
{\cal H}_{\rm EM}(p))
\]
where ${\cal H}_{\pm}(p)$ and ${\cal H}_{\rm EM}(p)$ are given
respectively by Eqs.~\ref{eq: odindva} and ~\ref{eq: odincet}.

As the Hamiltonian, the quantum momentum and boost generators 
are not uniquely determined. We can use this arbitrariness in order  
to make them invariant under both topologically trivial and
nontrivial gauge transformations. Acting in the same way as 
before in Section~\ref{sec: exoti}, we obtain the physical momentum 
and boost generators in the form
\begin{eqnarray*}
\hat{\rm P}_{\rm phys} & = & ({\cal H}_{+}^{\rm phys}(0) - 
{\cal H}_{-}^{\rm phys}(0)),\\
\hat{\rm K}_{\rm phys} & = & -i \frac{\rm L}{2\pi}
\sum_{\stackrel{p \in \cal Z}{p \neq 0}} \frac{(-1)^p}{p}
({\cal H}_{+}^{\rm phys}(p) + {\cal H}_{-}^{\rm phys}(p) +
{\cal H}_{\rm EM}^{\rm phys}(p) ),
\end{eqnarray*}
where
\begin{eqnarray*}
{\cal H}_{\pm}^{\rm phys}(p) & \equiv & \hbar \frac{\pi}{\rm L}
\sum_{\stackrel{q \in \cal Z}{q \neq (0;-p)}}
{\rho}_{\rm tot,\pm}(p+q) {\rho}_{\rm tot, \pm}(-q),\\
{\cal H}_{\pm}^{\rm phys}(0) & = & {\cal H}_{\pm}^{\rm phys}(p=0),
\end{eqnarray*}
and
\[
{\cal H}_{\rm EM}^{\rm phys}(p) = \frac{\hbar}{p} \frac{e_{+}}{2\pi}
{\rho}_{\rm tot}^{\rm N}(p) \frac{d}{db} + 
\frac{e_{+}^2{\rm L}}{8{\pi}^2} \sum_{\stackrel{q \in \cal Z}
{q \neq (0;-p)}} \frac{1}{q(q+p)} {\rho}_{\rm tot}^{\rm N}(p+q)
{\rho}_{\rm tot}^{\rm N}(-q).
\]

$2$. Let us now construct the algebra of the physical Hamiltonian,
momentum and boost generators. For ${\rm N} \neq \pm 1$,
the relativistic invariance is broken, so this algebra is not 
certainly a Poincare one. 

For the operators  ${\cal H}_{\pm}^{\rm phys}(p)$,
${\cal H}_{\rm EM}^{\rm phys}(p)$ and ${\cal H}_{\rm EM}^{\rm phys}
(0) = \hat{\rm H}_{\rm EM}^{\rm phys}$, we get the
following commutation relations
\[
[{\cal H}_{\pm}^{\rm phys}(n), {\cal H}_{\pm}^{\rm phys}(m)]_{-}=
\pm {\hbar} \frac{2\pi}{\rm L} (n-m) {\cal H}_{\pm}^{\rm phys}(n+m),
\]
\[
[{\cal H}_{+}^{\rm phys}(0) - {\cal H}_{-}^{\rm phys}(0),
{\cal H}_{\rm EM}^{\rm phys}(p)]_{-} = 
-{\hbar} \frac{2\pi}{\rm L} p
{\cal H}_{\rm EM}^{\rm phys}(p),
\]
and
\[
[{\cal H}_{+}^{\rm phys}(0) - {\cal H}_{-}^{\rm phys}(0),
{\cal H}_{\rm EM}^{\rm phys}(0)]_{-} =0,
\]
\[
[{\cal H}_{\pm}^{\rm phys}(p), {\cal H}_{\rm EM}^{\rm phys}(0)]_{-}=
[{\cal H}_{\pm}^{\rm phys}(0), {\cal H}_{\rm EM}^{\rm phys}(p)]_{-}.
\]
With these commutation relations ,
we easily obtain the algebra of the Poincare generators:
\[
[\hat{\rm H}_{\rm phys}, \hat{\rm P}_{\rm phys}]_{-}  =  0,  \\
\]
\[
[\hat{\rm P}_{\rm phys}, \hat{\rm K}_{\rm phys}]_{-}  =  -i\hbar
\hat{\rm H}_{\rm phys} + ({\rm boundary} \hspace{3 mm} {\rm terms})_1,
\]
and
\[
[\hat{\rm H}_{\rm phys}, \hat{\rm K}_{\rm phys}]_{-} =
-i\hbar \hat{\rm P}_{\rm phys} - i \frac{\rm L}{2\pi}
\sum_{\stackrel{p \in \cal Z}{p \neq 0}} \frac{(-1)^p}{p}
[{\cal H}_{\rm EM}^{\rm phys}(0), {\cal H}_{\rm EM}^{\rm phys}
(p)]_{-},\\
\]
\begin{equation}
+ ({\rm boundary} \hspace{3 mm} {\rm terms})_2,\\
\label{eq: pncr}
\end{equation}
where
\[
[{\cal H}_{\rm EM}^{\rm phys}(0), {\cal H}_{\rm EM}^{\rm phys}(p)]_
{-}=
-{\hbar}^3 \frac{e_{+}}{2\pi \rm L} \frac{1}{p} {\cal F}_{+-}(p)
{\rho}_{\rm tot}^{\rm N}(p) \frac{d}{db} 
\]
\[
+ {\hbar}^2
\frac{e_{+}^2}{4{\pi}^2} \sum_{\stackrel{q \in \cal Z}{q \neq (0;-p)}}
{\cal F}_{+-}(q) \frac{1}{q(p+q)} {\rho}_{\rm tot}^{\rm N}(p+q)
{\rho}_{\rm tot}^{\rm N}(-q),
\]
while the boundary terms are
\begin{eqnarray*}
({\rm boundary} \hspace{3 mm} {\rm  terms})_1 & \equiv & i
{\hbar}{\rm L}
(\hat{\cal H}_{+}^{\rm phys}(\frac{\rm L}{2}) +
\hat{\cal H}_{-}^{\rm phys}(\frac{\rm L}{2}) +
\hat{\cal H}_{\rm EM}^{\rm phys}(\frac{\rm L}{2})),\\
({\rm boundary} \hspace{3 mm} {\rm  terms})_2 & \equiv & i
{\hbar}{\rm L}
(\hat{\cal H}_{+}^{\rm phys}(\frac{\rm L}{2}) -
\hat{\cal H}_{-}^{\rm phys}(\frac{\rm L}{2})).\\
\end{eqnarray*}
The algebra obtained differs from the Poincare one for
, the difference being in the boundary terms and in the commutator
$[{\cal H}_{\rm EM}^{\rm phys}(0), {\cal H}_{\rm EM}^{\rm phys}
(p)]_{-}.$
The curvature ${\cal F}_{+-}$ associated with the projective
representation of the gauge group makes this commutator nonvanishing
for the models with anomaly.  This is another point where the
nonvanishing curvature ${\cal F}_{+-}$ shows itself (recall the
commutator of the modified momentum operators). 

For ${\rm N}=\pm 1$,  ${\cal F}_{+-}$ vanishes and 
up to the boundary terms we get the Poincare 
algebra. In the limit ${\rm L} \to \infty$, these boundary terms
vanish on the physical states, because the energy density is
assumed to diminish at spatial infinities faster than $1/{\rm L}$.
Otherwise, the total energy of the system would become infinite.
Therefore, for the ${\rm N} = \pm 1$ models on the line, the
Poincare algebra closes exactly. However, the boundary terms do
not affect the spectrum of the physical Hamiltonian which is
relativistic in the case of the circle, too.

For ${\rm N} \neq \pm 1$, in the limit ${\rm L} \rightarrow \infty$
the first term in the commutator $[{\cal H}_{\rm EM}^{\rm phys}(0),
{\cal H}_{\rm EM}^{\rm phys}(p)]_{-}$ disappears (since the gauge 
field 
has no global gauge-field degrees of freedom), while the second 
one survives. For the ${\rm N} \neq \pm 1$ models 
the Poincare algebra does not close even on the line. This means that 
such models are not relativistically invariant.

We can conclude that the nonclosure of the Poincare algebra in
~\ref{eq: pncr} is essentially due to the projective representation of 
the 
local gauge symmetry. Working on the circle allowed us  to construct 
explicitly the Poincare algebra breaking term connected to
the nonvanishing curvature ${\cal F}_{+-}$.

Let us
note that the Poincare algebra fails to close in the physical sector
where the states satisfy the quantum Gauss law constraint
~\ref{eq: dvanol} and the Poincare generators are gauge-invariant.
The physical Hamiltonian and momentum generator commute, so 
the translational
invariance is preserved. The origin of the breakdown of the 
relativistic
invariance lies in the anomaly or, more exactly, in the fact that
the local gauge symmetry is realized projectively.

\section{Charge Screening}
\label{sec: scree}
Let us introduce a pair of external charges, namely, a positive 
charge
with strength $q$ at $x_0$ and a negative one with the same strength
at $y_0$ \cite{iso90}. The external current density is
\[
j_{\rm ex,0}(x) = q(\delta(x-x_0) - \delta(x-y_0)) =
\frac{q}{\rm L} \sum_{p \in \cal Z} j_p^{\rm ex} 
e^{-i \frac{2\pi p}{\rm L}x},
\]
where
\[
j_p^{\rm ex} \equiv e^{i\frac{2\pi p}{\rm L}x_{0}} - 
e^{i\frac{2\pi p}{\rm L}y_0}.
\]
The total external charge is zero, so the external current density
has vanishing zero mode, $j_0^{\rm ex}=0$.
The Lagrangian density of the generalized CSM changes as follows
\[
{\cal L} \longrightarrow {\cal L} + A_0 \cdot j_{\rm ex,0}.
\]
The classical generalized CSM with the external charges 
added can be quantized
in the same way as that without external charges. The quantum 
Gauss law operator becomes
\[
\hat{\rm G}_{\rm ex} \equiv \hat{\rm G} + j_{\rm ex,0} =
\partial_1 \hat{\rm E} + e_{+} \hat{j}_{+} + 
e_{-} \hat{j}_{-} + j_{\rm ex,0} .
\]
Its Fourier expansion is
\[
\hat{\rm G}_{\rm ex} = \hat{\rm G}_0 +
\frac{2\pi}{{\rm L}^2} \sum_{p>0} (\hat{\rm G}_{+}^{\rm ex}(p)
e^{i\frac{2\pi}{\rm L}px} - \hat{\rm G}_{-}^{\rm ex}(p)
e^{-i\frac{2\pi}{\rm L}px} ),
\]
where
\begin{eqnarray*}
\hat{\rm G}_{+}^{\rm ex}(p) & \equiv & \hat{\rm G}_{+}(p) +
\frac{q\rm L}{2\pi} (j_p^{\rm ex})^{\star} , \\
\hat{\rm G}_{-}^{\rm ex}(p) & \equiv & \hat{\rm G}_{-}(p) -
\frac{q\rm L}{2\pi} j_p^{\rm ex}. 
\end{eqnarray*}
The physical states $|{\rm phys};A;{\rm ex} \rangle$ are defined as
\[
\hat{\tilde{\rm G}}_{\pm}^{\rm ex}(p) |{\rm phys};A;{\rm ex} \rangle 
\equiv (\hat{\rm G}_{\pm}^{\rm ex}(p) \pm  
\frac{1}{\hbar} \frac{e_{+}^2{\rm L}^2}
{8{\pi}^2} (1-{\rm N}^2)\alpha_{\pm p}) |{\rm phys};A;{\rm ex} 
\rangle =0.
\]
The physical quantum Hamiltonian becomes
\[
\hat{\rm H}_{\rm phys} = \frac{1}{2\rm L}\hat{\pi}_{b}^2 - 
\frac{1}{2} 
(\xi_{+} + \xi_{-}) + {\hbar} \frac{2\pi}{\rm L} \sum_{p > 0}   
({\rho}_{\rm tot,+}(-p) {\rho}_{\rm tot,+}(p)   +
{\rho}_{\rm tot,-}(p) {\rho}_{\rm tot,-}(-p)) + {\rm V}_{\rm ex},
\]
where
\[
{\rm V}_{\rm ex} \equiv  \frac{e_{+}^2{\rm L}}{8{\pi}^2}
\sum_{\stackrel{p \in \cal Z}{p \neq 0}} \frac{1}{p^2} 
({\rho}_{\rm N}^{\rm tot}(-p) +  \frac{q}{e_{+}}
j_{p}^{\rm ex}) \cdot
({\rho}_{\rm N}^{\rm tot}(p) +  \frac{q}{e_{+}}
(j_{p}^{\rm ex})^{\star} )
\]
is the Coulomb energy in the presence of the external charges.
Two new interactions contribute to the Coulomb energy owing to
the external charges. One is the classical Coulomb interaction
between the external charges and the other is the interaction between
the total internal current and the external current.

After some calculations we rewrite the physical Hamiltonian as
\[
\hat{\rm H}_{\rm phys} = \frac{1}{2\rm L} \hat{\pi}_{b}^2
-\frac{1}{2} (\xi_{+} +\xi_{-}) +
\sum_{p>0} \frac{{\rm E}_p}{p} \{ ( \overline{\rho}_{\rm tot,+}
(-p) + {\kappa}_{p,+} j_p^{\rm ex}) (\overline{\rho}_{\rm tot,+}(p) +
{\kappa}_{p,+} (j_p^{\rm ex})^{\star}) 
\]
\begin{equation}
+(\overline{\rho}_{\rm tot,-}(p) + {\kappa}_{p,-}(j_p^{\rm ex})
^{\star})
(\overline{\rho}_{\rm tot,-}(-p) + {\kappa}_{p,-}{j_p^{\rm ex}}) \}
+ {\hbar}^2 \frac{q^2}{\rm L} \sum_{p>0} \frac{1}{{\rm E}_p^2
({\rm N})}
j_p^{\rm ex} (j_p^{\rm ex})^{\star}
\label{eq: cetpet}
\end{equation}
with
\begin{eqnarray*}
{\kappa}_{p,+} & \equiv & \hbar \frac{qe_{+}}{2\pi} 
\frac{1}{{\rm E}_p^2({\rm N})}
(\cosh{t_p} + {\rm N} \sinh{t_p}), \\
{\kappa}_{p,-} & \equiv & \hbar \frac{qe_{+}}{2\pi} 
\frac{1}{{\rm E}_p^2({\rm N})}
(\sinh{t_p} + {\rm N} \cosh{t_p}).
\end{eqnarray*}
Comparing this Hamiltonian with the physical Hamiltonian without
the external charges, we see that the external charges change the
ground state. The ground state of the physical Hamiltonian
~\ref{eq: cetpet} satisfies
\begin{eqnarray*}
(\overline{\rho}_{\rm tot,+}(p) + {\kappa}_{p,+} \cdot
(j_p^{\rm ex})^{\star})
|{\rm ground};{\rm ex} \rangle & = & 0, \\
(\overline{\rho}_{\rm tot,-}(-p) + {\kappa}_{p,-} \cdot
j_p^{\rm ex})
|{\rm ground};{\rm ex} \rangle & = & 0,
\hspace{5 mm} p>0. \\
\end{eqnarray*}
The last term in ~\ref{eq: cetpet} is just the energy
of the ground state
\[
{\rm E}_0 = \langle {\rm ground};{\rm ex}| \hat{\rm H}_
{\rm phys} |{\rm ground};{\rm ex} \rangle =
{\hbar}^2 \frac{q^2}{\rm L} \sum_{p>0} \frac{1}{{\rm E}_p^2({\rm N})}
j_p^{\rm ex} (j_p^{\rm ex})^{\star}.
\]
The energy ${\rm E}_0$ depends only on the distance between the two
external charges:
\[
{\rm E}_0 =  {\hbar}^2 \frac{2q^2}{\rm L} \sum_{p>0}
\frac{1}{{\rm E}_p^2({\rm N})}
\{1 - \cos(\frac{2\pi p}{\rm L}(x_0 -y_0))\}
\]
\[
=\frac{q^2}{2{\rm M}_{\rm N}} \frac{\cosh{\frac{\rm L {\rm M}_{\rm N}}
{2}} -
\cosh(\frac{\rm L {\rm M}_{\rm N}}{2} - {\rm M}_{\rm N}|x_0 -y_0|)}
{\sinh{\frac{\rm L {\rm M}_{\rm N}}{2}}},
\]
where ${\rm M}_{\rm N}^2=\frac{e^2}{2\pi}\frac{1}{\hbar}(1+
{\rm N}^2)$.
In the limit $\rm L \gg 1$, we obtain
\[
{\rm E}_0 = \frac{q^2}{2{\rm M}_{\rm N}} (1 - e^{-{\rm M}_{\rm N}
|x_0 - y_0|}),
\]
i.e. the ground state energy has the form of the Yukawa type
potential. The long-range Coulomb force between widely separated
external charges disappears. Since there is no long-range force,
the external charges are screened. To show this we calculate
currents induced by the charges. The induced current (or charge
density) is
\[
\langle {\rm ground};{\rm ex}|:\hat{j}_{+}(x) + {\rm N} 
\hat{j}_{-}(x):
|{\rm ground};{\rm ex} \rangle  = - \frac{1}{2} {\rho}_{\rm bgrd} +
\varphi(x,x_0) - \varphi(x,y_0),
\]
where   $(- \frac{1}{2} {\rho}_{\rm bgrd})$ is a current induced
by the background charge, while $\varphi(x,x_0)$ and $\varphi(x,y_0)$
are currents induced by the external charges at points $x_0$ and
$y_0$ correspondingly,
\[
\varphi(x,y) \equiv - \sum_{n>0} \hbar \frac{e_{+}q}{\pi{\rm L}}
(1+{\rm N}^2) \frac{1}{{\rm E}_n^2({\rm N})} 
\cos(\frac{2\pi n}{\rm L}(x-y)) =
- \frac{q{\rm M}_{\rm N}}{2e_{+}} 
\frac{\cosh(\frac{\rm L {\rm M}_{\rm N}}{2} - {\rm M}_{\rm N}|x-y|)}
{\sinh{\frac{\rm L {\rm M}_{\rm N}}{2}}}.
\]
The current induced by the two external charges is a sum of the 
currents induced by each charge. 

In the limit $\rm L \gg 1$, 
\[
\varphi(x,x_0) \simeq - \frac{q{\rm M}_{\rm N}}{2e_{+}} 
e^{-{\rm M}_{\rm N}|x-x_0|}
\]
and damps exponentially as $x$ goes far from $x_0$ . 
Screening occurs only globally. The induced 
charge density distribution is
spread within the range of the order ${\rm M}_{\rm N}^{-1}$.
So if we are far away from the external charges , we can not find 
them.

The screening of each external charge occurs independently of the
other charges. That is, the charge density induced around any of the
two external charges does not depend on the location of the other one.
Next, the external charges are screened independently of the fact
whether the background charge vanishes or not.

The screening mechanism works therefore in the generalized CSM, too.
When a charge is placed in the system, the accompanying external
current polarizes the vacuum producing the complete compensation
of the charge. The background linearly rising electric field
characteristic for models with ${\rm N} \neq \pm 1$ does not
influence this mechanism.

\section{Discussion}
\label{sec: discu}
We have shown that the anomaly influences essentially the physical
quantum picture of the generalized CSM. 

$i)$ For the models with ${\rm N} \neq \pm 1$     and
defined on $S^1$, when
the gauge field has a global physical degree of freedom, the left--
right asymmetric matter content results in the background linearly 
rising electric field .This is a new physical effect caused
just by the anomaly and absent in the models without the anomaly,
i.e. for ${\rm N}=1$ (the standard Schwinger model) and
${\rm N}= -1$ (axial electrodynamics).

This effect distinguishes the generalized CSM on $S^1$ from 
the model defined
on $R^1$ as well. In the latter case, the gauge field has neither
local nor global physical degrees of freedom and the background field
disappears.

$ii)$ The anomaly leads also to the breakdown of the relativistic 
invariance. For the quantum theory of both ${\rm N}= \pm 1$ and
${\rm N} \neq \pm 1$ models we have presented the exotic statistics
matter formulation. In this formulation the physical Hamiltonian
is written in a compact diagonalized form. For the models with 
anomaly, the spectrum of the physical Hamiltonian turns out to be
non-relativistic and does not contain a massive boson. 

We have constructed the  physical quantum Poincare generators and 
shown that their algebra is not a Poincare one. 
We have demonstrated a relation between the anomaly, Berry phase
and breakdown of the relativistic invariance. Namely , the curvature
${\cal F}_{+-}$ related to the Berry phase does not vanish because
of the left-right asymmetric matter content. At the same time, just
the nonvanishing ${\cal F}_{+-}$ makes the algebra of the Poincare
generators different from the Poincare one.

The Poincare algebra fails to close on the physical states in the
chiral $QCD_2$ as well \cite{sara96}. The origin of the breakdown
of the relativistic invariance is the same   in both models and
lies in the anomaly. It would be of interest  to study the question
whether the relativistic invariance is broken for other models
with the projective realization of a local gauge symmetry, especially
in higher dimensions.

$iii)$  The total screening of charges characteristic for the SM
takes place in the generalized CSM, too. External charges are
screened globally even in the background linearly rising electric
field. The current density induced by the external charges damps
exponentially far away from them independently of the background
charge.

Due to Schwinger \cite{schw63}, the total screening of 
external charges implies the existence of a massive particle.
The breakdown of the relativistic invariance does not mean in
principle that in the anomalous models the dynamical mass generation
mechanism fails and that the massive particle can not exist.
Using the fact that the physical Hamiltonian and momentum commute,
we may try to prove the existence of their simultaneous eigenstates
of the relativistic massive particle energy-momentum relation
and then to identify these states with massive physical particles.
For the ${\rm N}=0$ model defined on $R^1$, such massive eigenstate
is constructed in \cite{sarad91}. The existence 
of the massive eigenstates for the ${\rm N} \neq \pm 1$ models
defined on $S^1$ will be investigated. We intend to report on that
in a future publication.


\begin{thebibliography}{44}
\bibitem{schw63} J. Schwinger, Phys.Rev. D {\bf 128},2425 (1962);\\
in {\it Theoretical Physics, Trieste Lectures}, 1962 (IAEA, Vienna,
1963 ).  
\bibitem{jack85} R. Jackiw and R. Rajaraman , Phys.Rev.Lett.
{\bf 54}, 1219 (1985).
\bibitem{raja85} R. Rajaraman, Phys.Lett. {\bf B154}, 305 (1985).
\bibitem{wign39} E. Wigner, Ann.Math. {\bf 40}, 149 (1939).
\bibitem{jack83} R. Jackiw, in {\it Relativity, Groups and
Topology II} \\
( Les Houches Summer School 1983) (North--Holland,
Amsterdam, 1984).
\bibitem{nels85} P. Nelson and L. Alvarez-Gaume, Commun.Math.Phys.
{\bf 99}, 103 (1985).
\bibitem{fadd86} L. D. Faddeev and S. L. Shatashvili, Phys.Lett.
{\bf B167}, 225 (1986).
\bibitem{fadd84} L. D. Faddeev, Phys.Lett. {\bf B14}, 81 (1984); \\
Nuffield Workshop on Supersymmetry and Supergravity, 1985.
\bibitem{hall86} I. G. Halliday, E. Rabinovici and A. Schwimmer,
Nucl.Phys.{\bf B268}, 413 (1986).
\bibitem{para88} M. B. Paranjape, Nucl.Phys.{\bf B307}, 649 (1988).
\bibitem{sarad91} F. M. Saradzhev, Int.J.Mod.Phys.{\bf A6},
3823 (1991);{\bf A8},2915, 2937 (1993).
\bibitem{niemi85} A. Niemi and G. Semenoff, Phys.Rev.Lett.
{\bf 55}, 927 (1985); {\bf 56}, 1019 (1986).
\bibitem{niemi86} A. Niemi and G. Semenoff, Phys.Lett. {\bf B175},
439 (1986).
\bibitem{semen87} G. Semenoff, in {\it Super Field Theory}, H. Lee
et al, eds. ( Plenum, NY, 1987).
\bibitem{sarad92} F. M. Saradzhev, Phys.Lett. {\bf B278}, 449
(1992).
\bibitem{sarad94} F. M. Saradzhev, Phys.Lett. {\bf B324}, 192
(1994).
\bibitem{dirac64} P. A. M. Dirac, {\it Lectures on Quantum Mechanics}
( Yeshiva Univ., NY , 1964).
\bibitem{mant85} N. S. Manton, Ann. Phys. {\bf 159}, 220 (1985).
\bibitem{raje88} S. Rajeev, Phys.Lett. {\bf B212}, 203 (1988).
\bibitem{hetr88} J. E. Hetrick and Y. Hosotani, Phys.Rev.
D {\bf 38}, 2621 (1988); \\
Phys.Lett. {\bf B230}, 88 (1989).
\bibitem{sarad88} F. M. Saradzhev, Sov.Phys. -- Lebedev Inst.
Reports, n. 9, 57 (1988); \\
Int.J.Mod.Phys. {\bf A9}, 3179 (1994); A. O. Barut and
F. M. Saradzhev,\\
Ann.Phys. (N.Y.) {\bf 234}, 220 (1994).
\bibitem{niese86} A. Niemi and G. Semenoff, Phys.Repts. {\bf 135},
99 (1986).
\bibitem{schiff68} L. I. Schiff, {\it Quantum Mechanics} (McGraw--
Hill, NY , 1968).
\bibitem{berry84} M. V. Berry, Proc.R.Soc. London, {\bf A392},
45 (1984).
\bibitem{jpha} F.M. Saradzhev , {\it "Berry Phase in Generalized
Chiral $QED_2$"},\\ 
hep-th/9609110, to appear in J.Phys. {\bf A}.
\bibitem{jack76} R. Jackiw and C. Rebbi, Phys.Rev.Lett.
{\bf 37}, 172 (1976).
\bibitem{callan76} C. G. Callan, Jr., R. Dashen and D. J. Gross,
Phys.Lett. {\bf B63}, 334 (1976).
\bibitem{iso90} S. Iso and H. Murayama, Progr.Theor.Phys.
{\bf 84}, 142 (1990).
\bibitem{sara96}  F.M. Saradzhev, Phys.Lett. {\bf B372}, 283 (1996). 
\end{thebibliography}
\end{document}